\RequirePackage{fix-cm}
\documentclass[natbib]{svjour3}                     
\smartqed  
\usepackage{graphicx}
\usepackage{amssymb}
\usepackage{amsmath}
\usepackage{subfigure}
\usepackage[usenames,dvipsnames]{color}
\usepackage{rotating}

\newcommand{\Sun}{\odot}
\newcommand{\lone}{$\mathcal{L}_1$\ }
\newcommand{\bp}[2]{$\mathrm{B}^{\scriptscriptstyle{\mathtt{#2}}}_{\scriptscriptstyle{\mathtt{#1}}}$}
\newcommand{\ud}{\mathrm{d}}
\newcommand{\ui}{\mathrm{i}}

\journalname{Celestial Mechanics and Dynamical Astronomy}

\begin{document}

\title{Evolution of the $\mathcal{L}_1$ halo family in the radial solar sail CRTBP}

\author{Patricia Verrier    \and
        Thomas Waters		\and
        Jan Sieber
}

\institute{P. Verrier \at
              Department of Mathematics, University of Portsmouth, Lion Gate Building, Lion Terrace, Portsmouth, Hampshire, PO1 3HF, UK\\
              \email{patricia.verrier@port.ac.uk}            \\
             \emph{Present address:} Advanced Space Concepts Laboratory, Department of Mechanical and Aerospace Engineering, University of Strathclyde, Glasgow G1 1XQ, UK\\
             \email{patricia.verrier@strath.ac.uk} 
           \and
           T. Waters \at
              Department of Mathematics, University of Portsmouth, Lion Gate Building, Lion Terrace, Portsmouth, Hampshire, PO1 3HF, UK              
           \and
           J. Sieber \at
           		CEMPS, University of Exeter, Exeter, EX4 4QF, UK
}

\date{Received: date / Accepted: date}

\maketitle

\begin{abstract}

We present a detailed investigation of the dramatic changes that occur in the \lone halo family when radiation pressure is introduced into the Sun-Earth circular restricted three-body problem (CRTBP). This photo-gravitational CRTBP can be used to model the motion of a solar sail orientated perpendicular to the Sun-line. The problem is then parameterized by the sail lightness number, the ratio of solar radiation pressure acceleration to solar gravitational acceleration. Using boundary-value problem numerical continuation methods and the AUTO software package \citep{Doedel1991} the families can be fully mapped out as the parameter $\beta$ is increased. Interestingly, the emergence of a branch point in the retrograde satellite family around the Earth at $\beta\approx0.0387$ acts to split the halo family into two new families. As radiation pressure is further increased one of these new families subsequently merges with another non-planar family at $\beta\approx0.289$, resulting in a third new family. The linear stability of the families changes rapidly at low values of $\beta$, with several small regions of neutral stability appearing and disappearing. By using existing methods within AUTO to continue branch points and period-doubling bifurcations, and deriving a new boundary-value problem formulation to continue the folds and Krein collisions, we track bifurcations and changes in the linear stability of the families in the parameter $\beta$ and provide a comprehensive overview of the halo family in the presence of radiation pressure. The results demonstrate that even at small values of $\beta$ there is significant difference to the classical CRTBP, providing opportunity for novel solar sail trajectories. Further, we also find that the branch points between families in the solar sail CRTBP provide a simple means of generating certain families in the classical case.


\end{abstract}

\section{Introduction}
\label{intro}

Solar sail space craft use radiation pressure on a reflective sail that can be oriented relative to the Sun-line for propulsion. With the recent success of the IKAROS interplanetary solar sail mission \citep{ikaros} and the upcoming launch of the Sunjammer mission such craft have become a reality. The additional acceleration such space craft experience changes their orbital dynamics and permits novel trajectories and concepts. In this work we investigate how the halo family of periodic orbits near the $\mathcal{L}_1$ Lagrangian point of the classical circular restricted three-body problem (CRTBP) changes when radiation pressure is introduced.  

The CRTBP has long been shown to be useful for the development of space craft missions. Periodic orbits near $\mathcal{L}_1$, particularly the halo family, have been used in missions such as ISEE-3, SOHO and Genesis \citep{Gomez2001, Koon1999}. For a solar sail the $\mathcal{L}_1$  equilibrium point can be moved closer to the Sun, providing advantages for solar observation. The Geostorm mission is one such proposed solar sail mission aimed at providing early warnings of solar storms \citep{West2004}. Understanding how the dynamics around the $\mathcal{L}_1$ point change for a solar sail allows the design of such missions to target specific characteristics, such finding as orbits with particular periods, closest approaches to the Sun or certain distances from the Earth. The study of the periodic orbits is the first step in gaining an extensive overview of the solar sail dynamics around the displaced Lagrangian point.

The efficiency of a solar sail can be parameterized through the lightness number $\beta$, which is the ratio of the solar radiation pressure acceleration to the solar gravitational acceleration. In a recent review of solar sail technology, \citet{Macdonald2011} discuss examples of near-term, mid-term and far-time solar sail missions. These are respectively: GeoSail, a mission to place a solar sail in the Earth's magnetotail; the Solar Polar Orbiter, a mission to place a solar sail in a polar orbit around the Sun; and the Interstellar Heliopause Probe, a mission to measure the interstellar medium. From the characteristic accelerations provided, and assuming a sail efficiency of 85\% the lightness numbers of these missions can be calculated (following \citealt{McInnes1999}) as $\beta\approx0.02$ for GeoStorm, $\beta\approx0.1$ for the Solar Polar Orbiter and $\beta\approx0.3-0.6$ for the Interstellar Heliopause Probe. For contrast, using values for the sail area and total mass given in \citet{ikaros}, IKAROS has $\beta\approx0.001$ while the upcoming Sunjammer mission has $\beta$ in the range 0.0388 to 0.0455 \citep{sunjammer}. Low values of the lightness number are thus of most interest to near-term applications. Note that in these calculations we have used a realistic value of 85\% efficiency for the sail material to convert the given physical parameters of the craft to a value of $\beta$ that can be compared to the theoretical model of a perfect sail which we will use in this paper. The theoretical model assumes the sail is ideal i.e. the acceleration is completely directed in the radial direction and sunlight is `perfectly' reflected.

The search for periodic orbits in the classical CRTBP is a large and active area of research. 
As the CRTBP is Hamiltonian with one first integral (the energy) periodic orbits generically come in one-parameter families characterised by the energy level for any given mass ratio. This is a well known property of Hamiltonian systems and of the CRTBP in particular, see e.g. \citet{Wintner1931, Deprit1969}. The simplest model of a solar sail in the Earth-Sun system is an extension of the CRTBP that includes radiation pressure directed in a radial direction (i.e. along the Sun-line). This problem is also Hamiltonian and as such it can be treated similarly to the CRTBP. We will call this modified problem the radial solar sail CRTBP, or RSCRTBP.

This problem was first studied in the astronomical context of binary star systems, where it is referred to as the photo-gravitational problem, for example in studies of the modified Lagrangian points by \citet{Chernikov1970}, \citet{Simmons1985} and \citet{Schuerman1980}. This early work showed that the Lagrangian points still exist in the solar sail problem, as well as one-parameter families of periodic orbits similar to the classical CRTBP. Recently, the same model has been used to investigate periodic orbits for solar sails by \citet{Farres2010}, \citet{Baoyin2006} and \citet{McInnes2000}. 

\citet{Baoyin2006} approximates individual halo orbits using a third-order analytical expansion at the $\mathcal{L}_1$ point in the RSCRTBP, looking at values of the sail lightness number $\beta$ up to about $0.3$. The authors find that the amplitude of these halo orbits increases as $\beta$ increases. 

\citet{McInnes2000} looks at how the $\mathcal{L}_1$ and $\mathcal{L}_2$ halo families in the Earth-Sun case change for seven values of $\beta$ between $0$ and $0.06$, using differential correction starting from an approximate analytical solution. The $\mathcal{L}_1$ halo family is seen to change in shape dramatically, while the $\mathcal{L}_2$ family merely decreases in amplitude. Partial linear stability information is also determined, and the $\mathcal{L}_1$ halo family is seen to have a large region of linear stability at the higher values of $\beta$. 

\citet{Farres2010} use multiple shooting methods to look at families of periodic orbits near to $\mathcal{L}_1$ and $\mathcal{L}_2$ for $\beta=0.051689$ (a lightness number proposed for the GeoStorm solar sail mission), and find that the associated planar and vertical Lyapunov families and the halo families still exist near the $\mathcal{L}_1$ point in this regime.

While the RSCRTBP is a simple extension for solar sails that are perpendicular to the sun-line it is not the only extension. In general the sail orientation relative to the sun-line can be described by two angles, and the problem is no longer Hamiltonian. However families of periodic orbits can be generated using either of the two angles as a parameter. If the model of the sail alignment results in a reversible system then natural families will also exist for a given angle as for the Hamiltonian case \citep{Devaney1976, Sevryuk}. \citet{Farres2010} look at the more general reversible system where the sail is parameterized by an angle $\delta$ to the sun-line (more specifically, one of the two angles is held constant so that the sail is perpendicular to the sun-line at $\delta=0$ but rotates in the plane defined by the $z$ direction and the solar radial vector as $\delta$ changes), which reduces to the RSCRTBP when $\delta=0$. When $\delta\neq 0$ the spatial $z \to -z$ symmetry of the problem is destroyed and the pitchfork bifurcation of the halo family from the planar Lyapunov family disappears. The planar Lyapunov family instead leaves the plane, splits and merges with the two branches of the halo family.

In the non-Hamiltonian versions of the solar sail problem some sail angles make orbits possible that are always out of plane. This has potential practical advantages, for example, one can generate `pole sitter' orbits sit over one of the Earth's poles for their entire orbit (e.g. \citealt{Waters2007}). There is also the possibility of extending the problem to the eccentric restricted three-body problem e.g \citet{Biggs2009, Biggs2009b}.

In this work we perform a comprehensive and detailed investigation of how the entire \lone halo family, and certain connected families, change with the lightness number $\beta$ in the RSCRTBP. To achieve this we use the boundary-value problem numerical continuation methods implemented in the AUTO software package \citep{Doedel1991,Doedel2011}. AUTO is a powerful tool for studying families of periodic orbits, as demonstrated by \citet{Doedel2007} in an extensive study of certain families of periodic orbits in the classical CRTBP. Here, we look at the linear stability and bifurcations of the halo family in the RSCRTBP over a large range of the lightness parameter, in most cases $0 \leq \beta \leq 0.5$, extending the results of \citet{McInnes2000} and investigating further the drastic change in shape of the halo family. In addition, we fully map out the linear stability of the family across a continuous range of $\beta$. Although some values of $\beta$ considered are not currently realistically obtainable they fit within the far-term mission examples discussed by \citet{Macdonald2011} and are included to gain a complete overview of the halo family in the RSCRTBP and enable the design of new solar sail missions. 

The main results of our paper are summarized at the beginning of Section \ref{sec:results}. In particular, Fig. \ref{fig:stabdiag} presents a complete qualitative bifurcation diagram of the RSCRTBP halo family in the two-parameter
plane. The remainder of Section \ref{sec:results} goes into the details of shape
changes for the orbit families shown in Fig. \ref{fig:stabdiag}. Linear stability
results are presented in Section \ref{sec:stability}.

Prior to presenting our results we introduce the necessary background information on model and methods: Section \ref{sec:model} describes the RSCRTBP geometry and mathematical model, Section \ref{sec:numcont} describes the numerical continuation schemes. Section \ref{sec:periodicorbits} describes the \lone halo family and its context in the classical Sun-Earth CRTBP, which serves as the starting point of our analysis.

\section{The radial solar sail circular restricted three-body problem (RSCRTBP)}
\label{sec:model}

The classical circular restricted three-body problem (CRTBP) considers the motion of a massless test particle under the influence of two point masses $m_1$ and $m_2$ in a circular orbit. It is often used to model the motion of a small body in planetary systems, for example either an asteroid in the Sun-Jupiter system or a space craft in the Earth-Moon system. A rotating reference frame with origin at the system center of mass is defined so that $m_1$ and $m_2$ remain fixed at $(-\mu,0,0)^T$ and $(1-\mu,0,0)^T$ respectively, where $\mu=m_2/(m_1+m_2)$ is the mass ratio. Units are used such that the distance between $m_1$ and $m_2$, their angular velocity and $G(m_1+m_2)$ are all unity, where $G$ is the gravitational constant. In this frame of reference the position of the test particle is $\boldsymbol{r}=(x,y,z)^T$ and 
\begin{eqnarray}
\boldsymbol{r}_1 &=& (x+\mu,y,z)^T\\
\boldsymbol{r}_2 &=& (x-1+\mu,y,z)^T
\end{eqnarray}
are its positions relative to the primary and secondary mass respectively.

\begin{figure}
\centering
\includegraphics[width=0.5\textwidth]{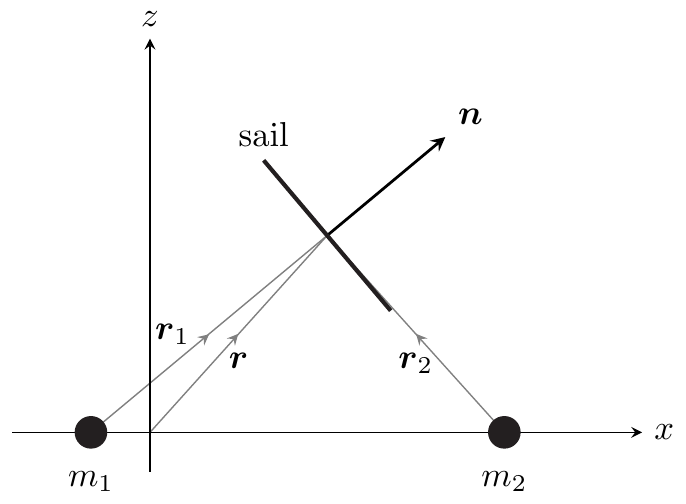}
\caption{\label{fig:sail} The location and orientation of the solar sail in the RSCRTBP, shown in the $x$-$z$ plane}
\end{figure}

To model the dynamics of a solar sail in the framework of the CRTBP the effects of radiation pressure on the test particle need to be included in the equations of motion. \citet{McInnes1999} models this additional acceleration on a perfectly reflecting solar sail as
\begin{equation}
\mathbf{a}=\beta\frac{GM_\Sun}{r_\Sun^2}(\hat{\boldsymbol{r}}_\Sun\cdot\boldsymbol{n})^2\boldsymbol{n}
\end{equation}
where $M_\Sun$ is the Sun's mass, $\boldsymbol{r}_\Sun$ is the vector between the sail and Sun and $r_\Sun=|\boldsymbol{r}_\Sun|$. In the context of the CRTBP the Sun is assumed to be the primary mass and we have $M_\Sun=m_1$ and $\boldsymbol{r}_\Sun=\boldsymbol{r}_1$. The parameter $\beta$ is the lightness number, defined as the ratio of the solar radiation pressure acceleration to solar gravitational acceleration. The classical case with no radiation pressure is obtained when $\beta=0$, and a value of $\beta=1$ would have radiation pressure equal to the gravitational force due to the Sun. As discussed typical values of $\beta$ for solar sails are fairly low, generally below 0.1. The vector $\boldsymbol{n}$ is the sail normal and defines the sail's alignment relative to the Sun, as illustrated in Fig. \ref{fig:sail}.

In this work we consider the case where the sail is always orientated along the Sun-line and the acceleration on the sail is in the radial direction only i.e. $\boldsymbol{n}=\boldsymbol{\hat{r}}_\Sun=\boldsymbol{\hat{r}}_1$. In this case 
\begin{equation}
\boldsymbol{a}=\beta\frac{(1-\mu)}{r_1^3} \boldsymbol{r}_1
\end{equation}
where $r_1=|\boldsymbol{r}_1|$ and the equations of motion of the RSCRTBP are given by
\begin{equation}
\ddot{\boldsymbol{r}}+2\boldsymbol{\omega}\times\dot{\boldsymbol{r}} = \nabla V
\label{eq:eom}
\end{equation}
where $\boldsymbol{\omega}=(0,0,1)^T$ is the rotation axis of the synodic frame and the potential $V$ is
\begin{equation}
V = \left(\frac{(1-\mu)(1-\beta)}{r_1}+\frac{\mu}{r_2} \right) + \frac{1}{2}|\boldsymbol{\omega}\times\boldsymbol{r}|^2 
\end{equation}
where $r_2=|\boldsymbol{r}_2|$. This new system differs from the CRTBP only by the modification to the $m_1$ gravitational potential term and is still Hamiltonian. The energy constant of the system is
\begin{equation}
E^\prime=\frac{1}{2}\left(\dot{x}^2+\dot{y}^2+\dot{z}^2\right) -\frac{1}{2}\left(x^2+y^2\right)-\frac{(1-\mu)(1-\beta)}{r_1}-\frac{\mu}{r_2}
\end{equation}
and a modified Jacobi constant can be defined as $C_J^\prime=-2E^\prime$. When $\beta=0$ the classical Jacobi constant $C_J$ is recovered.

\begin{figure}[!t]
\centering
\includegraphics[width=0.49\textwidth]{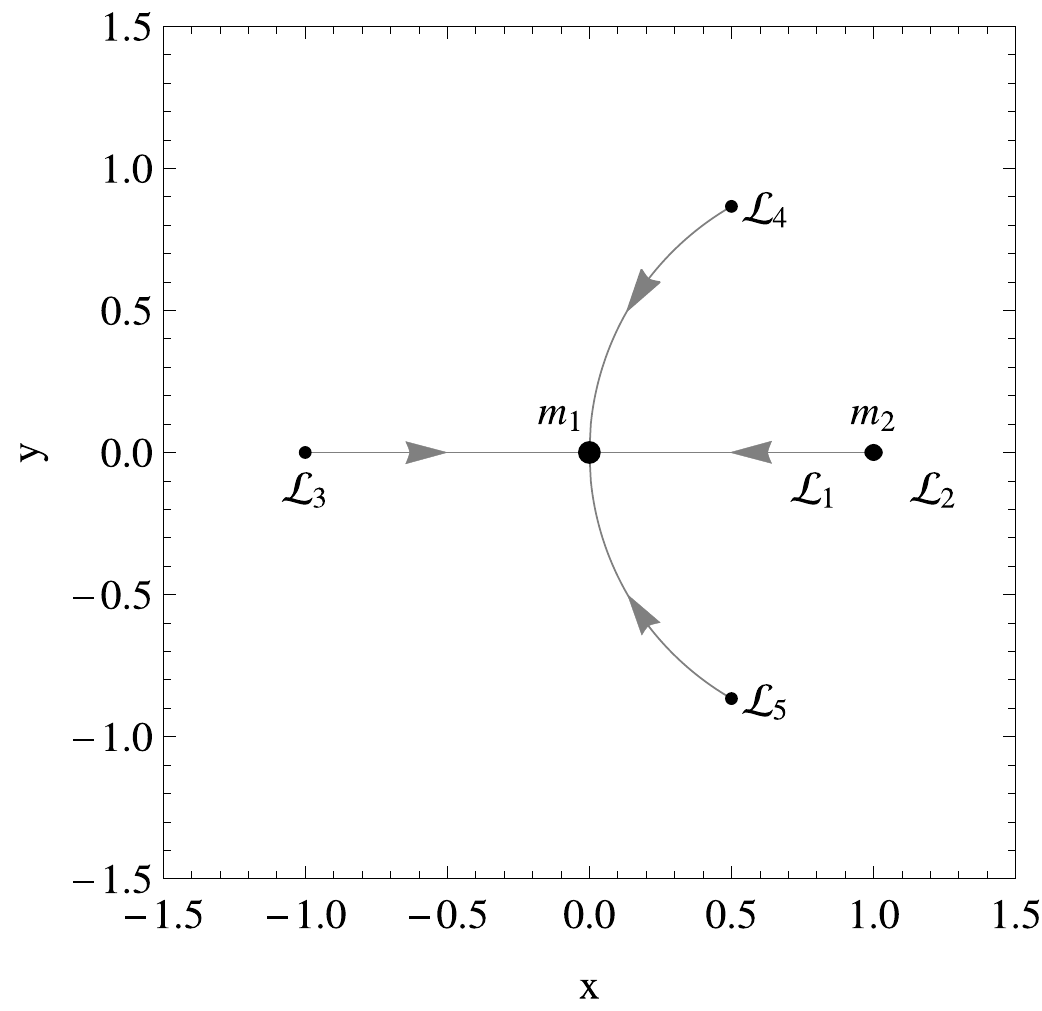}
\includegraphics[width=0.49\textwidth]{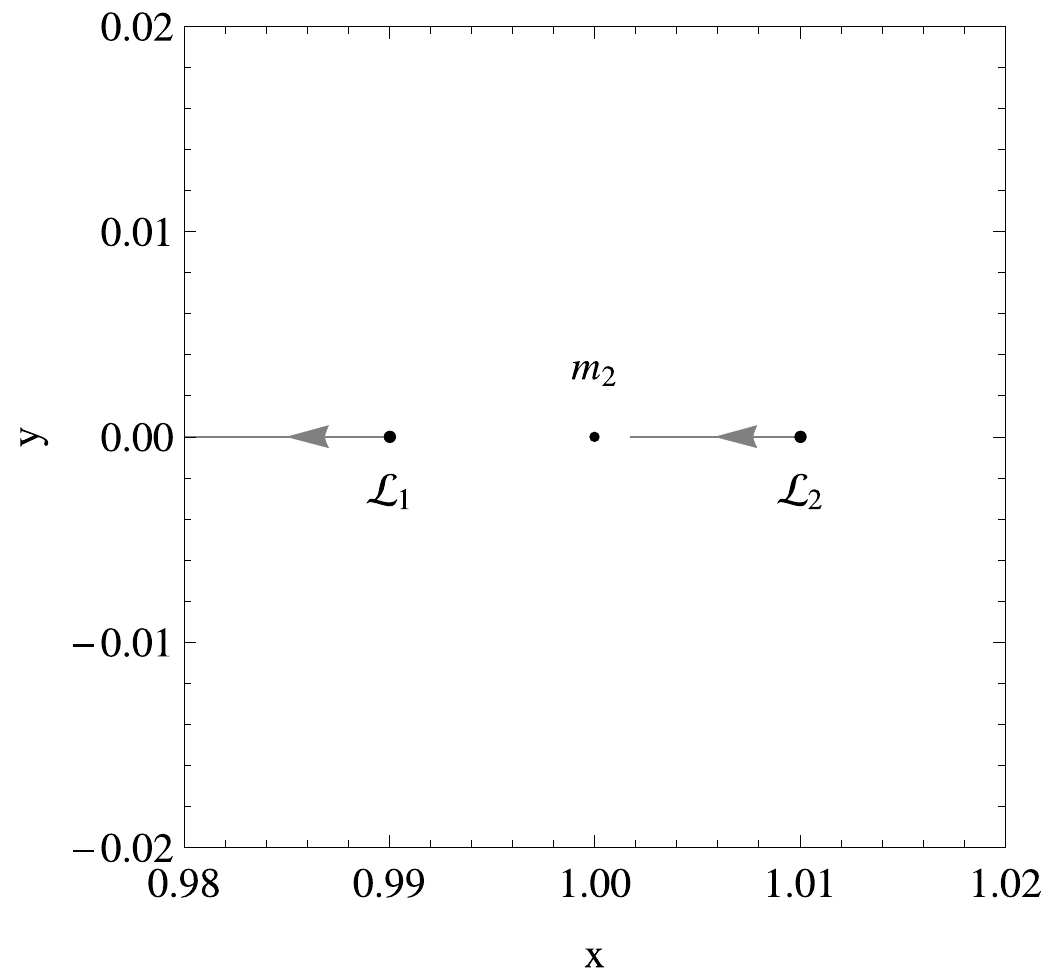}
\caption{\label{fig:Lpointssail}The change in position of the Lagrangian points in the RSCRTBP for the Sun-Earth mass ratio of $\mu=3 \times 10^{-6}$ for $\beta$ increasing from 0 to 1. A close-up of the near-Earth region is shown in the right-hand panel. The classical location of each equilibrium is labeled, and the locus in $\beta$ shown for each point}
\end{figure}

For $\beta < 1$ the five Lagrangian points persist, although their location is slightly modified depending on the value of the lightness parameter, as shown in Fig. \ref{fig:Lpointssail}. All the equilibria remain in the $x-y$ plane, and the collinear points remain on the $x$-axis. As $\beta$ tends to 1 the triangular points and $\mathcal{L}_1$ and $\mathcal{L}_3$ coalesce at the position of $m_1$, while $\mathcal{L}_2$ moves inwards towards $m_2$ \citep{Simmons1985}. The linear stability of the collinear points $\mathcal{L}_1$, $\mathcal{L}_2$, and $\mathcal{L}_3$ remains of type center-center-saddle as $\beta$ increases. The linear stability of the triangular points is also similar to the classical case, being center-center-center and (in general) linearly stable below a modified Routh ratio 
\begin{equation}
{\mu_R}_\beta = \frac{1}{2} \left(1 - \sqrt{\frac{32-9(1-\beta)^{2/3}}{36-9(1-\beta)^{2/3}}} \right) 
\end{equation}
and (an unstable) center-focus-focus above it i.e. with eigenvalues of the form $\pm \ui a, \pm b \pm \ui c$ with $a,b,c \in \mathbb{R}^{+}$ \citep{Simmons1985,Chernikov1970,Schuerman1980}. The modification is small, for example at $\beta=0.9$ the critical mass ratio is approximately ${\mu_R}_\beta\approx0.0302$ compared to $\mu_R\approx0.0385$ in the classical case.

As the RSCRTBP is still Hamiltonian with one first integral $E^\prime$ there are still one-parameter natural families of periodic orbits emanating from the Lagrangian points as in the classical case. As discussed, these orbits have been studied in the literature, both in the context of solar sails and astronomical models of radiation pressure \citep{Farres2010,Baoyin2006,McInnes2000}.

The equations of motion of the RSCRTBP, like the CRTBP, are invariant under the two transformations $(x,y,z,\dot x, \dot y, \dot z, t) \to (x, -y, z, -\dot x, \dot y, -\dot z, -t)$ and $(x,y,z,\dot x, \dot y, \dot z, t) \to (x, y,-z, \dot x, \dot y, -\dot z, t)$. Thus periodic orbits can be symmetric with respect to the $x$-$z$ plane (with time reversal), the $x$-$y$ plane, or both \citep{Gomez2001}. Halo orbits in particular are symmetric with respect to the $x$-$z$ plane.


\section{Numerical continuation}
\label{sec:numcont}

There are three main methods for generating periodic orbits: single shooting (often referred to as differential correction in celestial mechanics), multiple shooting and collocation.

Single shooting methods generate a periodic orbit by correcting the initial state until a fixed point of the return map is found. Multiple shooting methods are similar, but split the orbit up into segments. Collocation methods on the other hand discretize the periodic orbit as piecewise polynomials. Each of these discretization methods reduces the problem of finding a periodic orbit to numerically solving a finite-dimensional non-linear system of equations. Continuation of orbit families in system parameters can then be wrapped around any of these methods. In general, shooting methods are  simple to implement. Collocation methods are computationally more expensive but more robust when computing periodic orbits with rapidly varying properties (such as stability or shape) along the trajectory (see e.g. \citealt{Ascher1998}). As such they are well suited to dealing with orbits in the CRTBP that have close approaches with either of the primaries.

Historically, periodic orbits in the classical CRTBP were generated using single shooting, often incorporating methods to exploit the symmetry of some orbits, for example the method of \citet{Breakwell1979} to generate halo orbits. Early work on periodic orbits for solar sails also use such methods e.g. \citet{McInnes2000}, whereas more recent work has also used multiple shooting e.g. \citet{Farres2010}. Collocation methods have also become popular recently for both the classical CRTBP and solar sail periodic orbits, e.g. \citet{Ozimek2010, Doedel2003, Doedel2005, Doedel2007, Calleja2012}. For a recent review of the methods commonly used to generate solar sail trajectories see \citet{Wawrzyniak}. 

Outside of celestial mechanics the same methods are employed for numerous dynamical systems, with multiple shooting and collocation methods being commonly used. There are many standard software packages available. One of these is AUTO, 
which continues families of periodic orbits by formulating them as an extended boundary-value problem (BVP), discretizing the BVP using polynomial collocation with adaptive meshes and using pseudo-arclength continuation \citep{Doedel1981, Doedel1991, Doedel2011}. It is widely used and regarded as one of the most advanced continuation software packages available \citep{Krauskopf}. 

A number of recent studies use AUTO to investigate periodic orbits in the classical CRBTP. \citet{Doedel2007} investigate connections between families emanating from the five Lagrangian points in the CRTBP for a wide range of mass ratios, and determine the ranges over which these connections exist by continuing the branch points themselves. \citet{Calleja2012} use AUTO to calculate invariant manifolds of halo orbits, among others. The capability to not only generate and continue periodic orbits, but to automatically detect and continue branch points in a parameter such as the mass ratio makes AUTO a powerful tool for the analysis of dynamical systems such as the CRTBP. 

The halo family in the classical Earth-Sun problem experiences a very close approach with the Earth. Several other families relevant to the investigation here also suffer from close approaches and collisions with both masses. To accurately generate periodic orbits near close approaches a method such as collocation as implemented in AUTO is required. Although an alternative would be a regularization of the equations of motion to remove the singularities (see e.g. \citealt{szebehely}), in practice this introduces more complications, especially in the three-dimensional case. Since collocation methods are sufficient for our investigation of the RSCRTBP, regularization methods are not implemented. Note also that we use the term `collision' in the sense that the trajectory passes through the singularity at the location of either point mass. Close approaches that in reality would pass beneath the surface of the Earth (or Sun) are not detected in this work. Although such orbits individually are not of practical interest, the ability to continue past them allows other parts of the family that do not have collisions to be investigated, and permits new orbits to be found.

Our aim is to investigate the evolution of the halo family and connected families, as well as their bifurcations, for a wide range of $\beta$ values in the RSCRTBP. To do so we need to continue not only a large number of families, but the bifurcations within them, accurately and efficiently. As such AUTO is an obvious and very suitable choice for the method to do so. The next subsections describe the method for continuing periodic orbits in the RSCRTBP in AUTO in more detail.

\subsection{Continuation of families with AUTO}
\label{subsec:auto}

As discussed above, families of periodic orbits can be continued by formulating them as an extended boundary-value problem (BVP), discretizing the BVP using polynomial collocation and using pseudo-arclength continuation, as implemented in the AUTO package \citep{Doedel2011,Doedel1991}. The BVP equations must be provided to AUTO directly, and details of the application of this method for periodic orbits in the CRTBP are given in \citet{Doedel2007}, \citet{Doedel2003}, \citet{Doedel2005} and \citet{Calleja2012}, and will be summarized here. For the first order system given by
\begin{equation}
\dot{\boldsymbol{x}}(t) = f(\boldsymbol{x}(t),p)
\end{equation}
where for the RSCRTBP $\boldsymbol{x}(t) = (\mathbf{r},\dot{\mathbf{r}})\in \mathbb{R}^6$ and the parameters $p$ are $\mu$ and $\beta$, the general boundary-value problem defining a periodic orbit is given by
\begin{eqnarray}
\dot{\boldsymbol{x}}(t) &=& T f(\boldsymbol{x}(t),p) \label{eq:bvp1}\\ 
0 &=& \boldsymbol{x}(0) - \boldsymbol{x}(1) \label{eq:bvp2}\\
0 &=& \int_0^1 \dot{\boldsymbol{x}}_\mathrm{ref}^T \boldsymbol{x}(t) \ud t \label{eq:bvp3}
\end{eqnarray}
where $T$ is the (unknown) period of the periodic orbit and $\boldsymbol{x}_\mathrm{ref}$ is a nearby reference orbit (usually the solution previously found along the branch). Equation~\eqref{eq:bvp1} is the equation of motion, where time has been rescaled to fit the periodic orbit into the unit time interval $[0,1]$. Equation~\eqref{eq:bvp2} sets the periodic boundary conditions (in our case $6$-dimensional). Equation~\eqref{eq:bvp3} fixes the phase of the periodic orbit. If $\boldsymbol{x}(t)$ is a solution of \eqref{eq:bvp1} and \eqref{eq:bvp2} then so is $\boldsymbol{x}(t+s)$ for any shift $s$. Condition \eqref{eq:bvp3} determines the shift such that the new solution $\boldsymbol{x}$ has minimal distance from the reference orbit $\boldsymbol{x}_\mathrm{ref}$. To continue a family of periodic orbits one of the parameters (one component of $p$) is left free and the additional pseudo-arclength condition 
\begin{equation}\label{eq:parc}
s = \int_0^1 \boldsymbol{x}_\mathrm{tan}^T\left( \boldsymbol{x}(t) - \boldsymbol{x}_\mathrm{ref}(t) \right) \ud t + (T-T_\mathrm{ref}) T_\mathrm{tan} + (p-p_\mathrm{ref}) p_\mathrm{tan}
\end{equation}
is included, where $s$ is the \emph{pseudo-arclength}. In \eqref{eq:parc} $(\boldsymbol{x}_\mathrm{ref}(\cdot),T_\mathrm{ref},p_\mathrm{ref})$ is the previous periodic orbit in the family, $(\boldsymbol{x}_\mathrm{tan},T_\mathrm{tan},p_\mathrm{tan})$ is the approximate tangent to the branch in the previous point $(\boldsymbol{x}_\mathrm{ref}(\cdot),T_\mathrm{ref},p_\mathrm{ref})$, and $s$ is the approximate (desired) distance of the new solution $(\boldsymbol{x}(\cdot),T,p)$ from the previous solution given by $(\boldsymbol{x}_\mathrm{ref}(\cdot),T_\mathrm{ref},p_\mathrm{ref})$.
In the CRTBP the natural families of periodic orbits are parameterized by the energy level. However, the energy does not appear directly in the equations of motion as they are given in Eq. \ref{eq:eom}, so the continuation scheme described above cannot be immediately implemented. An alternative to reformulating the system that is also well suited to numerical implementation is provided by \citet{Munoz2003}, who introduce an artificial parameter $\Lambda$ into the equations of motion together with an `unfolding term'. \citet{Munoz2003} show that for a Hamiltonian system given by $\dot{ \boldsymbol{x}}=g_0(\boldsymbol{x})$ with a single first integral $I(\boldsymbol{x})$ the system
\begin{equation}\label{eq:unfold}
g(\boldsymbol{x},\Lambda) = g_0(\boldsymbol{x}) + \Lambda \nabla I(\boldsymbol{x})
\end{equation}
has a locally unique one-dimensional branch of periodic orbits with $\Lambda\equiv0$. These branches are parameterized by the value of the integral $I$ and can be continued in the unfolding parameter $\Lambda$ as there are no nearby periodic orbits for $\Lambda\neq0$. This method has been used by \citet{Doedel2003}, \citet{Doedel2005}, \citet{Doedel2007} and \citet{Calleja2012} to follow one-parameter families of orbits in the CRTBP, and is equally applicable to the RSCRTBP. As already discussed, \citet{Doedel2007} demonstrate how it can be used in AUTO to generate all elementary families associated with the Lagrangian points for a wide range of mass ratios. \citet{Calleja2012} also use it in AUTO to generate unstable manifolds of vertical and halo orbits in the CRTBP. The first integral in the CRTBP is the energy, and \citet{Doedel2003} and \citet{Doedel2005} use an unfolding term of the form $\nabla \tilde E$ where 
\begin{equation}
\tilde E = \frac{1}{2} \left( \dot{x}^2 + \dot{y}^2 + \dot{z}^2 \right) - \frac{1}{2}\left(x^2+y^2\right)-\frac{(1-\mu)}{r_1}-\frac{\mu}{r_2} - \frac{\mu(1-\mu)}{2}
\end{equation}
is related to the Jacobi constant by $C_J=-2\tilde{E}-\mu(1-\mu)$. However, \citet{Doedel2007} and \citet{Calleja2012} exploit the fact that the unfolding term in \eqref{eq:unfold} does not have to be a multiple of the gradient of first integral, $\nabla I(x)$. It is sufficient to unfold with any term $\Lambda \mathbf{v}$ as long as $\int \mathbf{v}^T\nabla I(x) \ud t\neq0$ along all periodic orbits of the family. This is guaranteed for much simpler terms such as $\mathbf{v}=(0,0,0,\dot{x},\dot{y},\dot{z})$. This artificial damping term $\mathbf{v}$ is sufficient to ensure that the branch with $\Lambda=0$ is unique, and this is the unfolding term we will also use for the RSCRTBP. Thus, the extended equations of motion for the RSCRTBP are 
\begin{eqnarray}
\dot{x} &=& v_x \nonumber\\
\dot{y} &=& v_y \nonumber\\
\dot{z} &=& v_z \nonumber\\
\dot{v}_x &=& 2 v_y + x - (1-\mu)(1-\beta)\frac{(x+\mu)}{r_1^3} - \mu \frac{(x-1+\mu)}{r_2^3}) + \Lambda v_x\nonumber\\
\dot{v}_y &=& -2 v_x + y - (1-\mu)(1-\beta) \frac{y}{r_1^3} - \mu \frac{y}{r_2^3} + \Lambda v_y\nonumber\\
\dot{v}_z &=& -(1-\mu)(1-\beta)\frac{z}{r_1^3} - \mu \frac{z}{r_2^3} + \Lambda v_z \label{eq:unfolding}
\end{eqnarray}
and families can be continued in AUTO as described. The unfolding parameter $\Lambda$ will always be zero to numerical precision for all calculated orbits. For the CRTBP and RSCRTBP either an initial reference orbit can be used to start the continuation, or families can be started by branching off from the Lagrangian points directly. The latter method is most commonly used (e.g. \citealt{Doedel2007}) and will be used here also.

\subsection{Linear stability, Floquet multipliers and bifurcations}
\label{subsec:floquet}

The linear stability of a periodic orbit is relevant to station-keeping strategies and the existence and computation of unstable and stable manifolds. For a periodic orbit in a Hamiltonian system such as the RSCRTBP linear stability is determined by the Floquet multipliers, the eigenvalues of the monodromy matrix (equivalent to the state transition matrix evaluated around the orbit, or the eigenvalues of the the linearized Poincar\'e map). \citet{Howell1999} provide an excellent overview of the application of Floquet theory to the CRTBP, which is also applicable to the RSCRBP. Briefly, the monodromy matrix is a real and  symplectic map so its eigenvalues are both reciprocal pairs and complex conjugate pairs. Further one pair of Floquet multipliers is equal to $+1$, corresponding to time shift and perturbations tangent to the family of periodic orbits. There are in general four different possibly configurations of the remaining two pairs of multipliers which correspond to the orders of instability as defined by \citealt{Howell1999}.
(As this is a Hamiltonian system it can be either neutrally stable or unstable, so the term `instability' will be used to refer to the linear stability.) These configurations are:
\begin{itemize}
\item order-0 instability (neutral stability at the linear level): both pairs of multipliers are on the unit circle;
\item order-1 instability: one pair is on the real axis and one pair is on the unit circle;
\item order-2 instability (real): both pairs are on the real axis, and
\item order-2 instability (complex): both pairs are in the complex plane and off the unit circle.
\end{itemize}
Note that neutral stability at the linear level does not in general imply neutral Lyapunov stability of the orbit in the full system.

The Floquet multipliers also provide information on the bifurcations that occur along a family, which are often associated with a change in the linear stability. Changes in linear stability occur generically through collisions of a pair of multipliers at $+1$ or $-1$ or collisions of two pairs of multipliers on the unit circle or real axis. A collision of a pair of multipliers at $+1$ is a tangent bifurcation: either a fold in the energy level occurs or a new family branches off, usually in a pitchfork (or \emph{symmetry-breaking}) bifurcation in the RSCRTBP. We refer to this latter case as a \textit{branch point}. We use the notation \bp{A}{B} to refer to a branch point between family $\mathtt{A}$ and family $\mathtt{B}$, where $\mathtt{A}$ has the higher symmetry. A collision of a pair at $-1$ is a period-doubling bifurcation. At this point another family branches off with twice the period of the original family. Two pairs of multipliers colliding on the unit circle is a Krein collision (also known as a secondary Hamiltonian Hopf bifurcation, torus bifurcation or Neimark-Sacker bifurcation). Two pairs of multipliers colliding on the real axis is an inverse Krein collision (or a modified secondary Hamiltonian Hopf bifurcation). Inverse Krein collisions are not really bifurcations because they do not change the instability order (it remains $2$), although they do change the nature of the Floquet multipliers. 

The $\mathcal{L}_1$ halo family is well-known to possess every bifurcation discussed above, as well as all types of linear stability (e.g. \citealt{Howell1984}, \citealt{Breakwell1979}, \citealt{Gomez2001}). The aim of this work is to investigate how the \lone halo family and its linear stability changes compared to the classical case when radiation pressure is introduced i.e. when $\beta$ is increased. Since linear stability regions in the classical case are bounded by various bifurcations the detection and continuation of these in the parameter $\beta$ can be used to map out the linear stability as radiation pressure is increased. 

The Floquet multipliers are calculated by AUTO for each periodic orbit, and using the unfolding method described in Section \ref{subsec:auto}, branch points to other families can be detected and continued in $\beta$. (A complete description of the implementation of branch point continuation in the case of the CRTBP is given in \citealt{Doedel2007}). Period-doubling bifurcations can be found by using twice the period in the continuation and detecting them as branch points. However, AUTO currently has no methods of detecting or continuing Krein collisions and inverse Krein collisions. In addition, using the system as given in Equation \ref{eq:unfolding} the folds cannot be detected or continued either as the energy does not appear in the equations of motion directly. Instead, we must develop a new formulation of the system that permits this. In the case of folds, we achieve this by including the energy level directly, as discussed in Section \ref{subsec:folds} below. To detect and continue Krein collisions and inverse Krein collisions we derive an extended system of equations that describes the collision of the Floquet multipliers. This new methodology is presented in Section \ref{subsec:krein}. These two extensions to the boundary-value problem continuation methods provide a complete toolset to track changes of linear stability in any system parameter such as $\beta$.


\subsection{Detection and continuation of folds}
\label{subsec:folds}

In this section we present a new formulation of the BVP of the RSCRTBP that allows folds in the energy level of this system to be detected in AUTO. To do this the one-parameter families must be continued in the energy directly, rather than the non-physical parameter $\Lambda$ detailed in Section \ref{subsec:auto}. This can be achieved by using an unfolding term of the form $(\tilde E_0 - \tilde E^\prime) \nabla \tilde E^\prime$ instead, so that the extended equations of motion are now
\begin{equation}\label{eq:energy:added}
\dot{\boldsymbol{x}} = f(\boldsymbol{x},\mu,\beta) + \left(\tilde E^\prime_0 -\tilde E^\prime \right) \nabla \tilde E^\prime
\end{equation}
where $f(\boldsymbol{x},\mu,\beta)$ is the original system given in Equation \eqref{eq:eom} with $\boldsymbol{x}=(\boldsymbol{r},\dot{\boldsymbol{r}})$ and 
\begin{equation}
\tilde E^\prime(\boldsymbol{x},\mu,\beta) = \frac{1}{2}(\dot{x}^2+\dot{y}^2+\dot{z}^2) -\frac{1}{2}(x^2+y^2) - \frac{(1-\mu)(1-\beta)}{r_1} - \frac{\mu}{r_2}    - \frac{1}{2}\mu(1-\mu)
\end{equation}
is the energy associated with the system given by Equation \eqref{eq:eom}. Equation~\eqref{eq:energy:added} implies that along trajectories $\boldsymbol{x}(t)$ of \eqref{eq:energy:added} the quantity $\tilde E^\prime$ satisfies
\begin{displaymath}
  \frac{\ud}{\ud t}\tilde E^\prime(\boldsymbol{x}(t))=|\nabla \tilde E^\prime(\boldsymbol{x}(t))|^2 \left(\tilde E^\prime_0 -\tilde E^\prime(\boldsymbol{x}(t)) \right)\mbox{.}
\end{displaymath}
Since this relation prevents sign changes of $ \frac{\ud}{\ud t}\tilde E^\prime(\boldsymbol{x}(t))$, $\tilde{E}^\prime$ can be periodic only if it is constant. Thus, $\tilde E^\prime$ takes the constant value $\tilde E^\prime_0$ on every periodic orbit. Consequently, the term $(\tilde E^\prime_0 - \tilde E^\prime)$ will be zero along every periodic orbit of \eqref{eq:bvp1}  such that a periodic orbit of \eqref{eq:bvp1} with energy level $\tilde E^\prime_0$ will be a periodic orbit of \eqref{eq:energy:added} with system parameter $\tilde E^\prime_0$. Moreover, every periodic orbit of \eqref{eq:energy:added} is also a periodic orbit of \eqref{eq:bvp1}, and $\tilde E'_0$ (which is a system parameter in \eqref{eq:energy:added}) will be its energy level. Then the energy  $\tilde E^\prime_0$ of the original system \eqref{eq:eom} can be used as the continuation parameter, which allows folds in  $\tilde E^\prime_0$ to be detected and continued in AUTO similarly to branch points.

\subsection{Detection and continuation of Krein collisions}
\label{subsec:krein}

In this section we derive a new extended system of BVP equations that define a Krein collision. This system can then be implemented in AUTO, or similar continuation schemes. To detect a Krein collision in a system such as the RSCRTBP a constraint is needed to define the occurrence of two pairs of multipliers colliding. Such a constraint is given by
\begin{equation}
B= \frac{A^2}{4}+2
\end{equation}
where
\begin{eqnarray}
A &=& \lambda + \frac{1}{\lambda} + \mu + \frac{1}{\mu} \\ 
B &=& \left( \lambda + \frac{1}{\lambda} \right) \left( \mu + \frac{1}{\mu} \right) + 2 
\end{eqnarray}
where the non-trivial Floquet multipliers are $(\lambda, \frac{1}{\lambda}, \mu, \frac{1}{\mu})$ \citep{Howell1999,Howard1987}. Thus a simple method of detecting such collisions is to look for zeros of the function $B - \frac{A^2}{4} - 2$ along the family of periodic orbits. To continue a Krein collision, the original boundary-value problem given in Equations \ref{eq:bvp1} to \ref{eq:bvp3} for the system $\dot{\boldsymbol{x}} = f(\boldsymbol{x}(t),p)$ can be extended in general as follows. If $\boldsymbol{x} \in \mathbb{R}^n$, define two more $n$-dimensional complex linear boundary-value problems
\begin{eqnarray}
\dot{\boldsymbol{y}}(t) &=& T\partial_{\boldsymbol{x}} f(\boldsymbol{x}(t),p) \boldsymbol{y}(t) \label{eqn6a}\\
\dot{\boldsymbol{z}}(t) &=& T\partial_{\boldsymbol{x}} f(\boldsymbol{x}(t),p) \boldsymbol{z}(t) \label{eqn6}
\end{eqnarray}
where $\boldsymbol{y}(t),\boldsymbol{z}(t) \in \mathbb{C}^n$, with the boundary conditions
\begin{equation}
\left( \begin{array}{c}
\boldsymbol{0}\\
\boldsymbol{0} \end{array}\right) = \boldsymbol{M} 
\left( \begin{array}{c}
\boldsymbol{y}(0) \\
\boldsymbol{z}(0) \end{array}\right) - 
\left( \begin{array}{c}
\boldsymbol{y}(1) \\
\boldsymbol{z}(1) \end{array}\right) 
\label{eqn8}
\end{equation}
where $\boldsymbol{M}\in\mathbb{C}^{2\times2}$, and integral conditions
\begin{equation}
0 = \int_0^1 \left( 
\begin{array}{cc}
\boldsymbol{y}_\mathrm{ref}(t)^\mathrm{H}\boldsymbol{y}(t) & \boldsymbol{y}_\mathrm{ref}(t)^\mathrm{H}\boldsymbol{z}(t) \\ 
\boldsymbol{z}_\mathrm{ref}(t)^\mathrm{H}\boldsymbol{y}(t) & \boldsymbol{z}_\mathrm{ref}(t)^\mathrm{H}\boldsymbol{z}(t) \\ 
\end{array}
\right)\mathrm{d}t- \left ( \begin{array}{cc}
1 & 0\\
0 & 1 \end{array}\right )
\label{eqn9}
\end{equation}
where the notation $\boldsymbol{y}_\mathrm{ref}^H$ refers to the complex conjugate transpose. This provides $n+1$ real conditions (Equations \eqref{eq:bvp2} and \eqref{eq:bvp3}) and $2n+4$ complex conditions (Equations \eqref{eqn8} and \eqref{eqn9}) for the $n+3$ real variables $\boldsymbol{x}(0)\in\mathbb{R}^n$, $T\in\mathbb{R}$ and $p\in \mathbb{R}^2$, and the $2n+4$ complex variables $\boldsymbol{y}(0)\in\mathbb{C}^n$, $\boldsymbol{z}(0)\in \mathbb{C}^n$ and $\boldsymbol{M} \in \mathbb{C}^{2\times2}$. The system given by Equations \eqref{eqn6a} to \eqref{eqn9} tracks a basis $(\boldsymbol{y},\boldsymbol{z})$ of a complex two-dimensional eigenspace of the linearization of the original boundary-value problem \citep{Beyn2001}. Overall we have  $5n+9$ (real) conditions for $5n+11$ (real) unknowns. The extended system ensures that the eigenvalues of $\boldsymbol{M}$ are Floquet multipliers of $\boldsymbol{x}(t)$. Adding the condition $B=A^2/4 + 2$ expressed as
\begin{equation}
0 = \mathrm{Re} \left[ \left(1-\frac{1}{\mathrm{det} \boldsymbol{M}} \right)^2 \left(\mathrm{det} \boldsymbol{M} - \frac{(\mathrm{Tr}\boldsymbol{M})^2}{4}\right) \right] \label{eqn10}
\end{equation}
provides the final necessary condition, ensuring that the computation results in a curve in the ($\beta$--energy level) parameter plane.

This method does not check the Krein signature, so detects only that the eigenvalues collide, not that there is a change in linear stability. However, the Floquet multipliers of individual families has been checked to determine if this occurs. This type of BVP is one of the problem types supported by AUTO, however the complex generalized eigenvector pair $(\boldsymbol{y},\boldsymbol{z})$ are needed to start the continuation and must be found by other means. As such an implementation in Matlab of the above extended system in the combination with the collocation and pseudo-arclength continuation methods has been used. An alternative approach is to track sign changes in the function $B-A^2/4-2$ on the (two-dimensional) solution surface of the system (\ref{eq:bvp1})--(\ref{eq:bvp3}) defining the periodic orbits. In all parts where we have used both approaches the results agreed.


\section{Overview of relevant periodic orbits in the classical Sun-Earth CRTBP}
\label{sec:periodicorbits}

\begin{figure}
\centering
\includegraphics[width=0.95\textwidth]{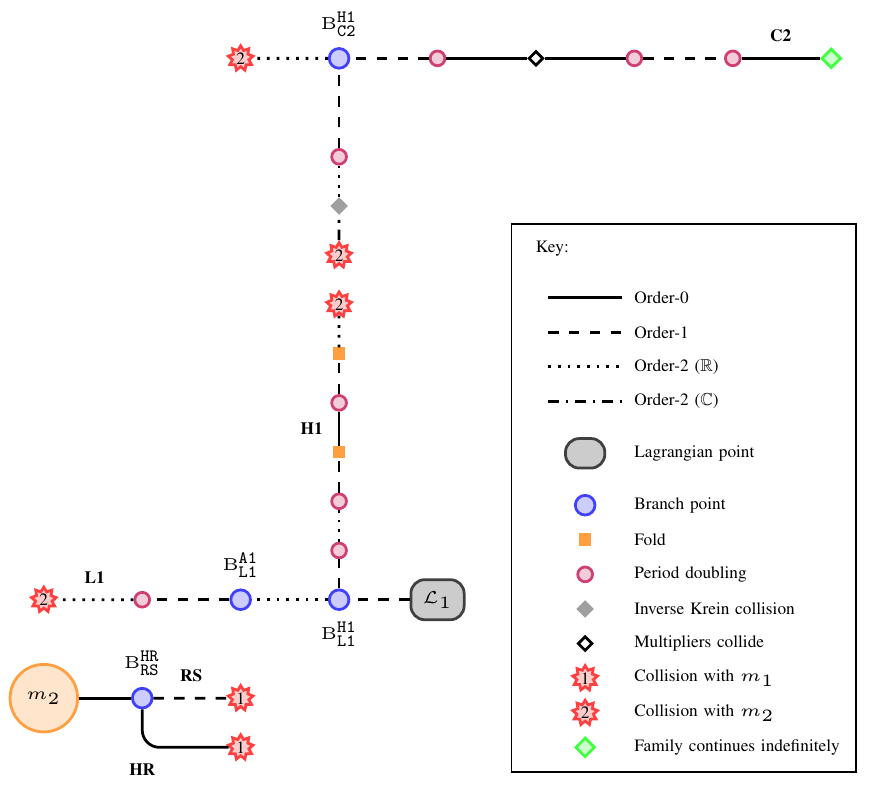}
\caption{\label{fig:bpdiag}Schematic bifurcation diagram and linear stability for selected families of periodic orbits in the Earth-Sun classical CRTBP near the secondary mass. Symbols are as defined in the key. The open black diamond in the C2 family marks the point where two pairs of multipliers pass through each other on unit circle}
\end{figure}

To provide a context for the evolution of the $\mathcal{L}_1$ halo family in the RSCRTBP we briefly summarise here the results for some relevant families in the classical Earth-Sun CRTBP. These families are the starting point for the evolution of the $\mathcal{L}_1$ halo family in the RSCRTBP. These are the planar Lyapunov family (L1), the $\mathcal{L}_1$ halo family (H1), a planar retrograde circular family around both masses (C2), the retrograde satellite (RS) family around $m_2$ (the Earth) and a non-planar bifurcating family from it (HR). We use the designations L1, H1 and C2 from \citet{Doedel2007} for the first three, and RS and HR for the two families around $m_2$. Fig. \ref{fig:bpdiag} shows the connections between these families and their bifurcations and linear stability, while portraits of each in physical space are shown in the Appendix. The RS and HR families are not connected to H1 in the classical Earth-Sun CRTBP, as shown in Fig. \ref{fig:bpdiag}, but they are included here as we find that they connect to H1 in the RSCRBP for certain values of $\beta$ (see Section \ref{sec:results}).

The well known planar Lyapunov family L1 is shown in Fig. \ref{fig:l1eseg} in the Appendix. This family contains two branch points, one to the halo family H1 and one to a family that \citet{Doedel2007} refers to as the Axial A1 family. The L1 family ends in a collision with $m_2$.

The halo family H1 branches from L1 and ends in a very close approach or collision with $m_2$. That is, the orbit becomes singular up to accuracy of the discretization. It is not clear in this case if this is a true collision with $m_2$ and this question does not seem to be addressed in the literature (the orbits at this point do however pass within the radius of the Earth, so there is little practical interest in this part of the family). As discussed in the previous section, at higher mass ratios this does not occur and the family continues, notably containing a branch point to a non-planar asymmetric family called W5 by \citet{Doedel2007} and the planar C2 family. The section of the family branching from C2 still exists in the Earth-Sun case, as shown in Fig. \ref{fig:bpdiag}. Both parts of the family in the Earth-Sun case are shown in Figs. \ref{fig:H1orbit_0} and \ref{fig:H1orbit_1} in the Appendix. Because the equations of motion of the CRTBP are invariant under the transformation $z\to-z$, a mirror image of this family under reflection in the $x-y$ plane also exists. The family shown in Figs. \ref{fig:H1orbit_0} and \ref{fig:H1orbit_1} is referred to as the north branch, and the reflected version the south branch. 

The C2 family is shown in Figure \ref{fig:c2eseg} in the Appendix. One side of this family ends in a collision with $m_2$, while the other appears to continue out to infinity. There are three period-doubling bifurcations seen near the start of this family, the middle of which is a connection to the $\mathcal{L}_2$ vertical Lyapunov family. 

The two families around $m_2$ shown in Fig. \ref{fig:bpdiag} are not considered in \citet{Doedel2007}. The first of these is a planar family usually referred to as the retrograde satellite family (e.g. \citealt{Henon1969}), labelled RS in Figure~\ref{fig:bpdiag}. The RS family corresponds to Str\"omgren's family \textit{f} \citep{Stroemgren1933}. It consists of roughly circular retrograde orbits about the secondary mass, as shown in Figure \ref{fig:rseseg} in the Appendix. The RS family ends in a collision with $m_1$ and, at this mass ratio, contains one branch point to a non-planar family with the same symmetry as H1 which we have called HR. 

The HR family is shown in Figure \ref{fig:hreseg} in the Appendix. Similar to the halo family it has both a north and south branch. It has linear stability of order-0 until it also ends in a collision with $m_1$. This family exists at higher mass ratios, but does not appear to be well known in the literature, so we also provide an example for the Earth-Moon mass ratio in Fig. \ref{fig:hrem} in the Appendix. In this case the collision with $m_1$ does not occur, and the family connects back to the RS family at a second branch point. It has order-0 instability for its entirety in this case as well.


\section{Periodic orbits in the Earth-Sun RSCRTBP}
\label{sec:results}

\subsection{Overview}

\begin{figure}[!hp]
\centering
\includegraphics[width=0.99\textwidth]{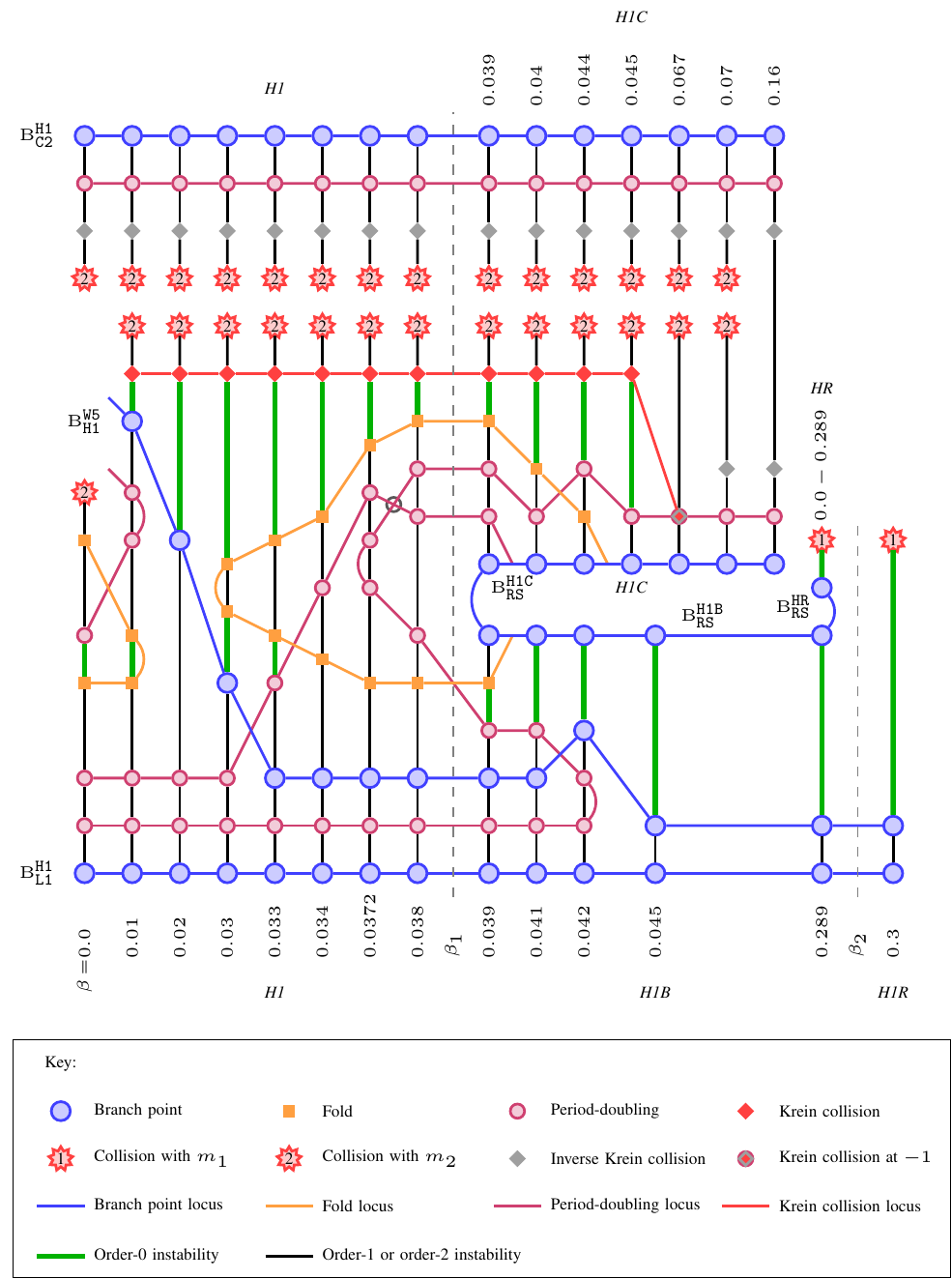}
\caption{\label{fig:stabdiag}
Schematic representation of the linear stability and bifurcations of the H1, HR, H1B, H1C and H1R families. Symbols are defined in the key. The vertical lines represent the families as labeled, and $\beta$ increases left to right. Green lines indicate order-0 instability. The H1 family always starts from the L1 branch point with order-1 instability, and the linear stability of the rest of the family can be deduced from the diagram. The gray dashed lines indicate the critical values $\beta_1$ and $\beta_2$ where the additional branch point in the RS family appears and disappears respectively. The position at which two period-doubling bifurcations merge briefly in H1 is marked with a gray circle, and the  point in H1C where the Krein collision occurs at $-1$ with a gray/red diamond inside a purple circle. Note that the H1C and HR1 families also exists beyond the maximum value of $\beta$ shown here}
\end{figure}

A summary of the results obtained for the evolution of the halo family in the Earth-Sun RSCRTBP for $\beta$ values up to 0.5 is shown in the topological bifurcation diagram in Fig. \ref{fig:stabdiag}. The vertical lines on this plot represent the halo family at different values of $\beta$, with the classical CRTBP being on the far left. The location of bifurcations causing changes in linear stability are marked, and the locus of each in $\beta$ is indicated. This is a schematic overview of the results from numerical continuation of both families and the bifurcations themselves, as described in Sec. \ref{sec:numcont}. 

The main result is a dramatic change in the halo family at $\beta\approx 0.039$, at which point the family connects to the retrograde satellite family via a new branch point. This change is accompanied by a gradual alteration of the shapes of the periodic orbits (as seen by \citet{McInnes2000} for the halo family for specific values of $\beta\leq0.05$). The H1 family is subsequently split into two separate families which have been labeled as H1B and H1C. At $\beta\approx0.289$ the H1B family merges with the HR family, to become what we have called the H1R family. From this point on there is little change in the H1C and H1R families as $\beta$ increases. The linear stability results also show significant changes occur in the H1 family even at low values of beta, and as we move along the family of periodic orbits for a given value of beta there is an intricate series of tangent bifurcations, period doublings, folds and Krein collisions, giving rise to the appearance and disappearance of regions of neutral linear stability (called order-0 instability in Sec. \ref{subsec:floquet}). The variation in the Floquet multipliers along a particular family (for fixed $\beta$) can be found by following the vertical lines and monitoring the sequence of bifurcations.

Representative plots of the families corresponding to Fig. \ref{fig:stabdiag} are shown in the Appendix. Fig. \ref{fig:egorbits1} shows the original H1 halo family and HR family at $\beta=0$. Fig. \ref{fig:H1orbit_03} shows H1 before the connection to the RS family, Fig. \ref{fig:H1Corbit_04} to \ref{fig:H1Corbit_5} show the two `new' families H1B and H1C that the original H1 family is split into for several values of $\beta$. Fig. \ref{fig:H1Rorbit_3} shows the H1R family formed by the merger of the H1B and HR families.

These results and the complex evolution of the halo family are discussed in greater detail in the following sections. To explain how the halo family changes with the sail lightness number $\beta$ in the Earth-Sun system it is first necessary to look at the evolution of the L1 and C2 families from which it branches. For a certain range of $\beta$ H1 also connects to the RS family, so we include a study of the evolution of this family as well. The evolution of all these families has been tracked for $\beta\leq1$.

\paragraph{Initialization of the computation of families}
The L1 family is found by branching from the $\mathcal{L}_1$ point, and the H1 family then found from the branch point in L1. The C2 family is found by starting from the branch point in the H1 family for a higher value of the mass ratio $\mu$ and then continuing it in the mass ratio $\mu$ to obtain the family in the Earth-Sun case. The RS family can be found by continuing C2 in $\beta$, as will be shown below. All other families studied in this section are connected to L1 and C2 via branch points, and thus found via branch switching in AUTO.

\subsection{The L1 family}

\begin{figure}[!b]
\centering
\includegraphics[width=0.6\textwidth]{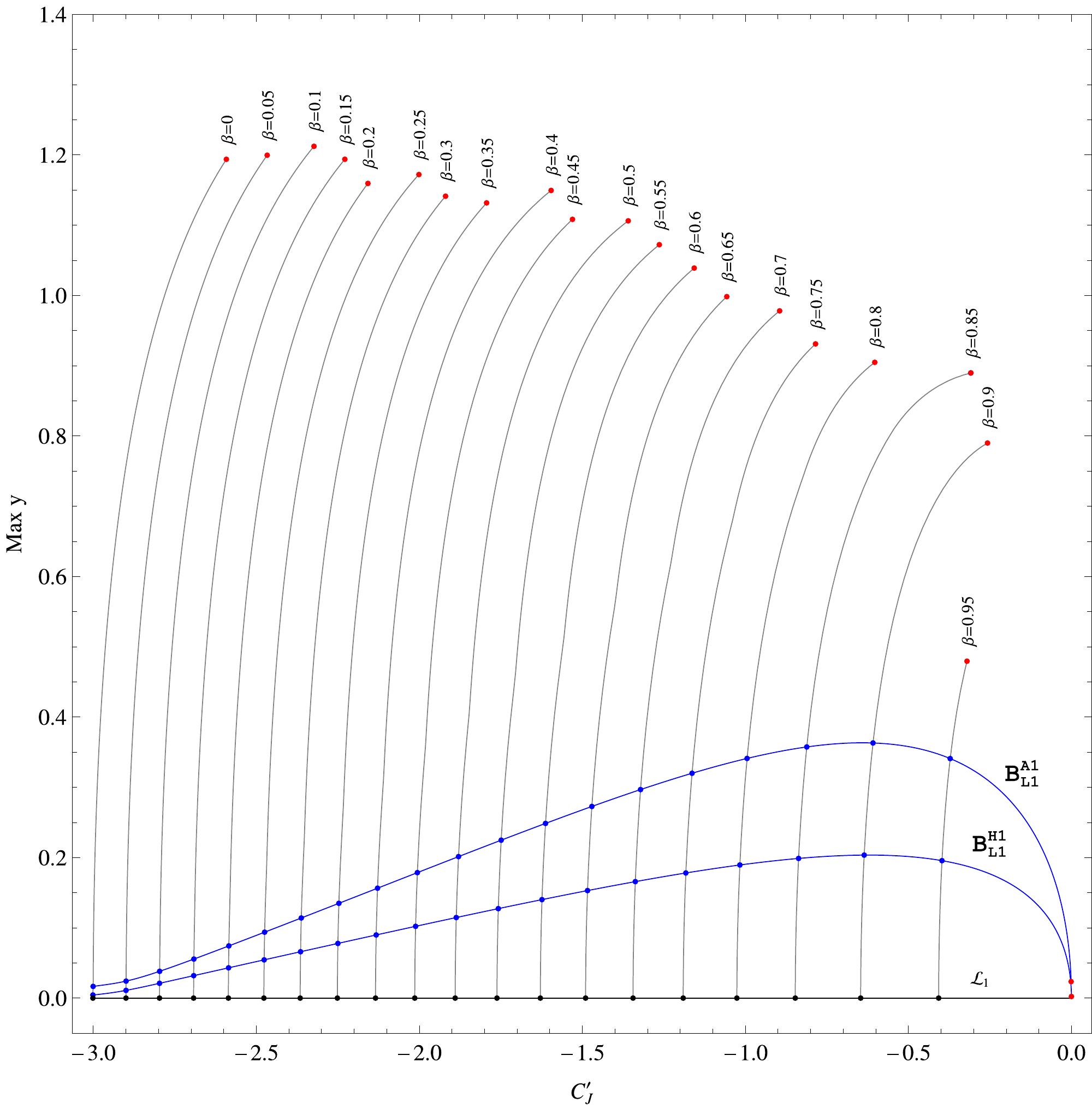}
\caption{\label{fig:l1tracks}Evolution of the L1 family for $0\leq\beta\leq0.95$. Each gray line represents an L1 family for a different value of $\beta$ as labeled. The locus of the \lone equilibrium is shown in black, and the H1 and A1 branch points in blue. Termination of the family due to collision is marked on each family as a red point}
\end{figure}

The L1 family remains qualitatively unchanged as the radiation pressure increases, with only the amplitudes of the orbits changing. Branch point continuation shows that the two branch points \bp{L1}{H1} and \bp{L1}{A1} to the H1 and A1 families exist up until almost $\beta=1$. In all cases the family still ends with a collision with $m_1$. Fig. \ref{fig:l1tracks} shows the maximum $y$ value of each orbit against its energy constant $C^\prime_J$ along the L1 family as a curve in the $y-C^\prime_J$ plane for varying $\beta$. This representation clearly illustrates the change in this planar family as radiation pressure increases. The evolution of the two branch points  in $\beta$, calculated using branch point continuation, is also shown in Fig.~\ref{fig:l1tracks}. As $\beta$ approaches 1 and the $\mathcal{L}_1$ point approaches the Sun these branching orbits both tend towards zero amplitude. Examples of the orbits in the L1 family for several values of $\beta$ are shown in Fig. \ref{fig:l1ev} in the Appendix.

\subsection{The C2 and RS families}

\begin{figure}[!t]
\centering
\includegraphics[width=0.65\textwidth]{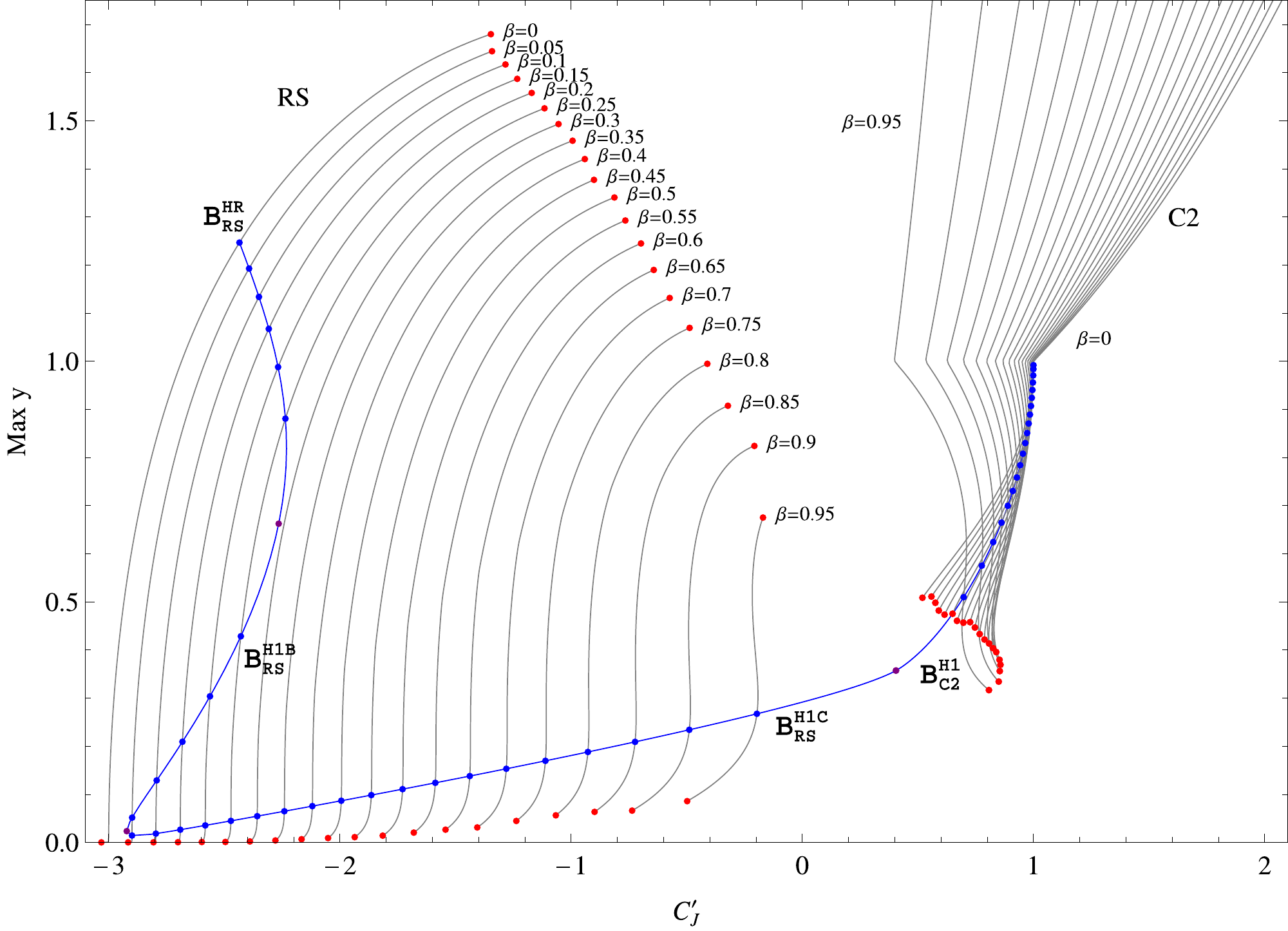}
\caption{\label{fig:c2rstracks} Evolution of the C2 and RS families for $0\leq\beta\leq0.95$. The C2 families are on the right and the RS families on the left. As before each gray line represents a single family for a different value of $\beta$, as labeled. For C2 only the two extremes have been labeled, but the same values have been used as for the RS family. Termination due to collision is marked on each family as a red point. The C2 families continue beyond the range of the plot. For the RS families the lower collision is with $m_2$ (the Earth) and the upper with $m_1$ (the Sun). (Note that the location of the $m_2$ collision is at a max $y$ value of 0 with $C^\prime_J$ at negative infinity, however the location shown on the graph is limited by numerical accuracy, as the orbits become elongated for high $\beta$ before converging to $m_2$.) The loci of the branch points in both families is plotted in blue, with folds in $\beta$ marked with a purple dot. The appearance of the two additional branch points on the RS family for certain $\beta$ values is clearly illustrated here}
\end{figure}

The C2 family also remains qualitatively unchanged as $\beta$ increases, as shown in the examples in Fig. \ref{fig:c2ev} in the Appendix. The evolution of the family is shown on the right side of Fig. \ref{fig:c2rstracks}. The locus of the branch point \bp{C2}{H1} is also shown. At $\beta=1$ this branch point merges with the \bp{RS}{HR} in the RS family, showing that the RS and C2 are a single family in the two-parameter plane $(\beta-C^\prime_J)$. The evolution of the RS family is shown on the left of Fig. \ref{fig:c2rstracks}, as well as the continuation of the branch point \bp{RS}{HR}. There are two folds in the energy level in the locus of this branch point at critical values of the lightness number of $\beta_1\approx0.0387$ and $\beta_2\approx0.289$. These are also marked in the figure. (It can be shown that one of the folds in $\beta$ is actually a cusp as it also corresponds to a fold in the energy level.) The result of these folds is the existence of two additional branch points along the RS family between the two critical values of the lightness number. As will be shown in the next section these two folds are linked to the evolution of the halo family. Example RS family orbits for three values of $\beta$ are shown in Fig. \ref{fig:rsev} in the Appendix.


\begin{sidewaysfigure}[!hp]
\vspace{5in}
\centering
\includegraphics[width=0.95\textwidth]{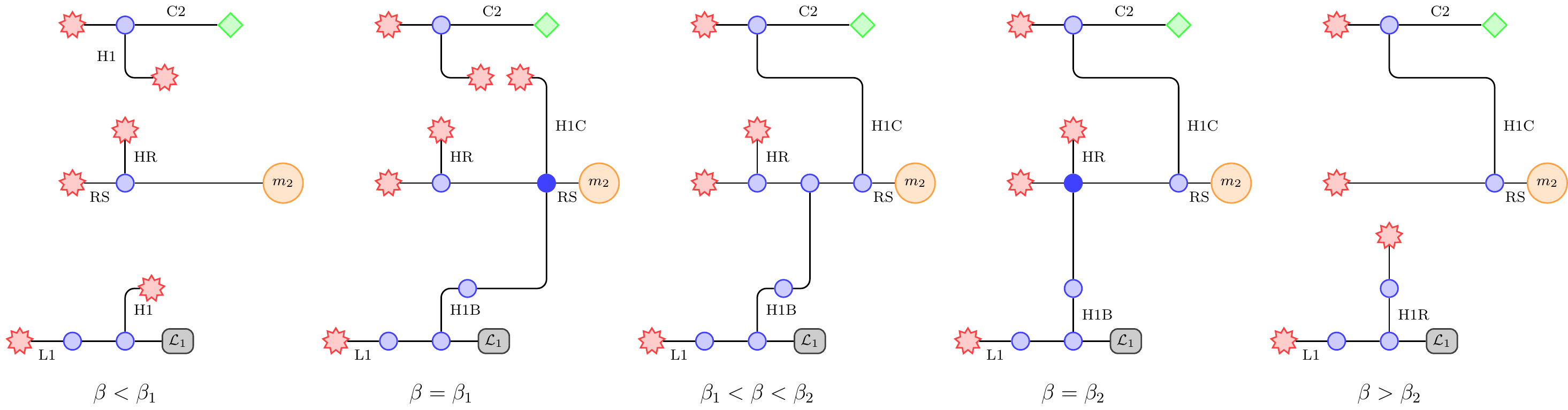}
\caption{\label{fig:bfdiag} 
Schematic bifurcation diagrams of H1 and related families for increasing values of $\beta$. Symbols are as before: blue circles represent branch points and red stars collisions. The $\mathcal{L}_1$ equilibrium point is shown as a grey rectangle and the secondary mass as a yellow circle. The green diamond indicates that the C2 family appears to continue out to infinity. The darker blue circles represent the two critical cases at $\beta_1$ and $\beta_2$ where two branch points in the RS family either appear or collide}
\end{sidewaysfigure}

\subsection{The H1 halo family}

The evolution of the H1 halo family in $\beta$ is very closely linked with the appearance and disappearance of the additional branch points in the RS family. As $\beta$ approaches the critical value $\beta_1$ the H1 family starts to fold back on itself spatially, down towards the ecliptic ($x-y$) plane. At $\beta=\beta_1$, the family intersects the $x-y$ plane and the planar orbit becomes the new branching orbit in the RS family. The original configuration for $\beta<\beta_1$ is shown in panel 1 of Fig. \ref{fig:bfdiag}, and this new configuration  at $\beta=\beta_1$ in panel 2.

\begin{figure}[!p]
\centering
\includegraphics[width=0.97\textwidth]{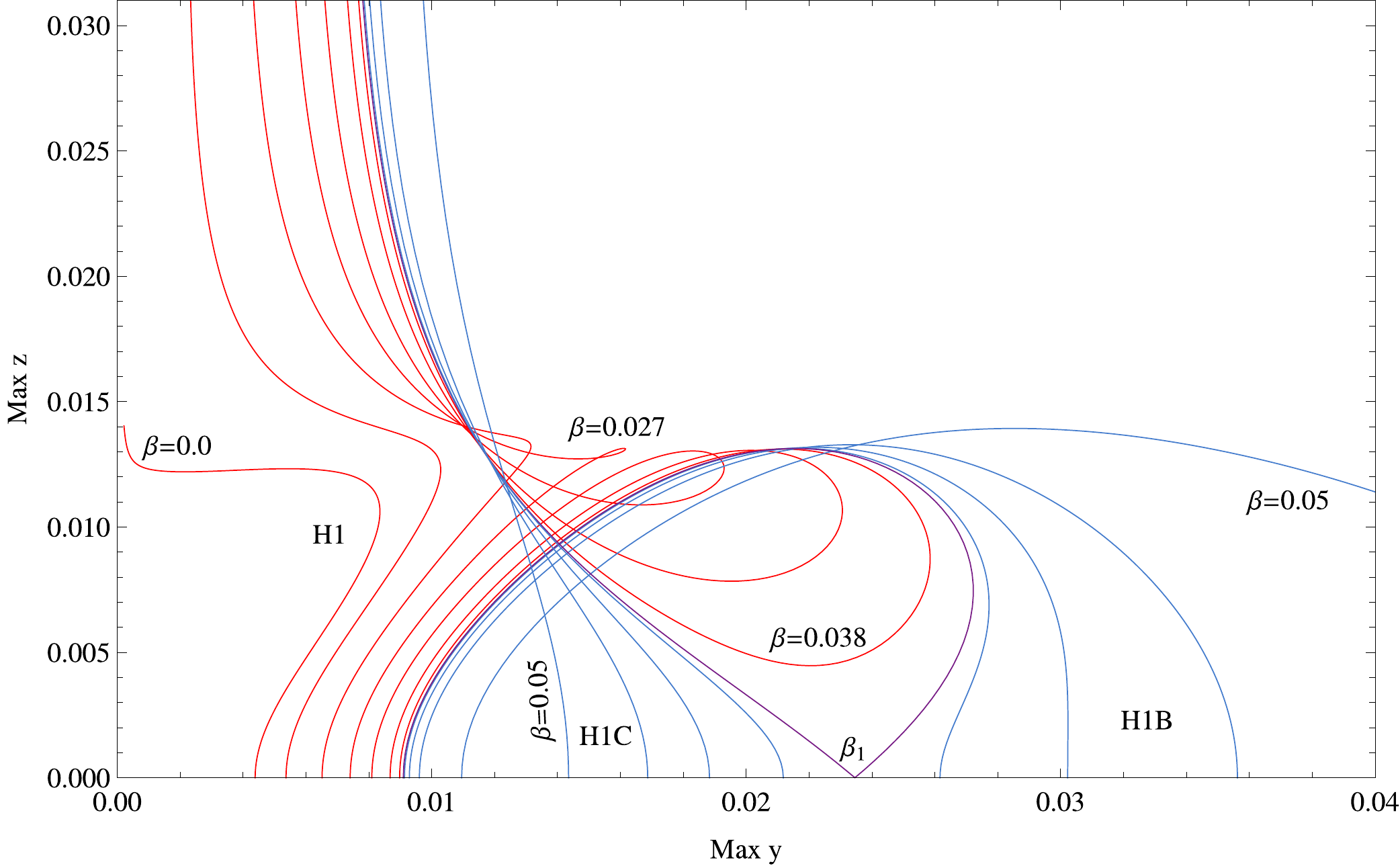}
\caption{\label{fig:bftracks1} Evolution of H1, H1B and H1C near $\beta_1$. Each line represents one family. The H1 families are shown in red for $\beta=(0.0,0.01,0.02,0.027,0.032,0.036,0.038)$ and the H1B and H1C families in blue for $\beta=(0.039,0.04,0.042,0.05)$. The critical case where H1 connects to the RS family at $\beta=\beta_1$ is shown in purple}
\end{figure}

The developing fold in the H1 family for $\beta<\beta_1$ is shown in the example family in Fig. \ref{fig:H1orbit_03}. The evolution is very clear in a plot of the tracks of the family in maximum $y$ and maximum $z$ position, as shown in red in Fig. \ref{fig:bftracks1}. This plot also shows the critical case  $\beta=\beta_1$ in purple, where the `loop' in the family touches the $x-y$ plane. 

As $\beta$ increases above $\beta_1$ and the new branch point in RS bifurcates into two that gradually move apart, the H1 family is split into two: a branch from the L1 family to the RS family, which we label H1B, and a branch from RS that ends in collision with $m_2$ as before. At $\beta\approx 0.16$ the amplitude of the orbits have changed sufficiently to avoid the collision and it connects to the C2 branch point. We label this family H1C. This configuration is shown in panel 3 of Fig. \ref{fig:bfdiag}. 

\begin{figure}[!bp]
\centering
\includegraphics[width=0.97\textwidth]{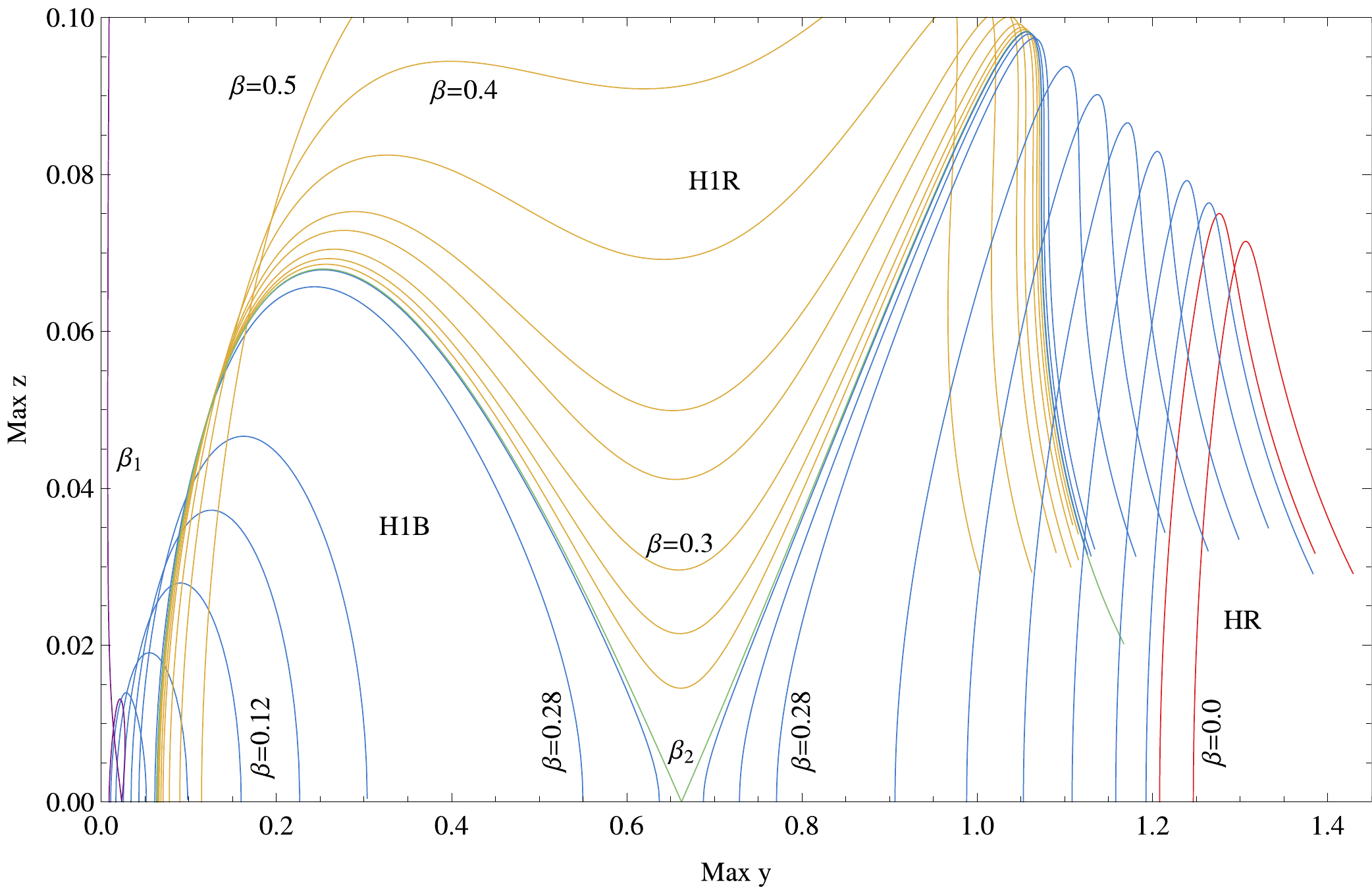}
\caption{\label{fig:bftracks2}
Evolution of H1B, HR and H1R in $\beta$. Each line represents one family. H1B families are on the left of the plot and HR on the right. Red lines are used for values of $\beta < \beta_1$, blue for $\beta_1<\beta<\beta_2$ and yellow for $\beta>\beta_2$. The critical case where H1 connects to the RS family at $\beta=\beta_1$ is shown in purple again, and the critical case where H1 connects to HR at $\beta=\beta_2$ is shown in green. H1B is shown for $\beta=(0.05,0.08,0.12,0.16,0.2,0.28,0.289)$, HR for $\beta=(0.0,0.036,0.05,0.08,0.12,0.16,0.2,0.24,0.28,0.286,0.289)$ and H1R for $\beta=(0.292,0.295,0.3,0.31,0.32,0.35,0.4,0.5)$. The abrupt end to the HR and H1R tracks is due to collision with $m_1$
}
\end{figure}

\begin{figure}[tp]
\centering
\includegraphics[width=0.9\textwidth]{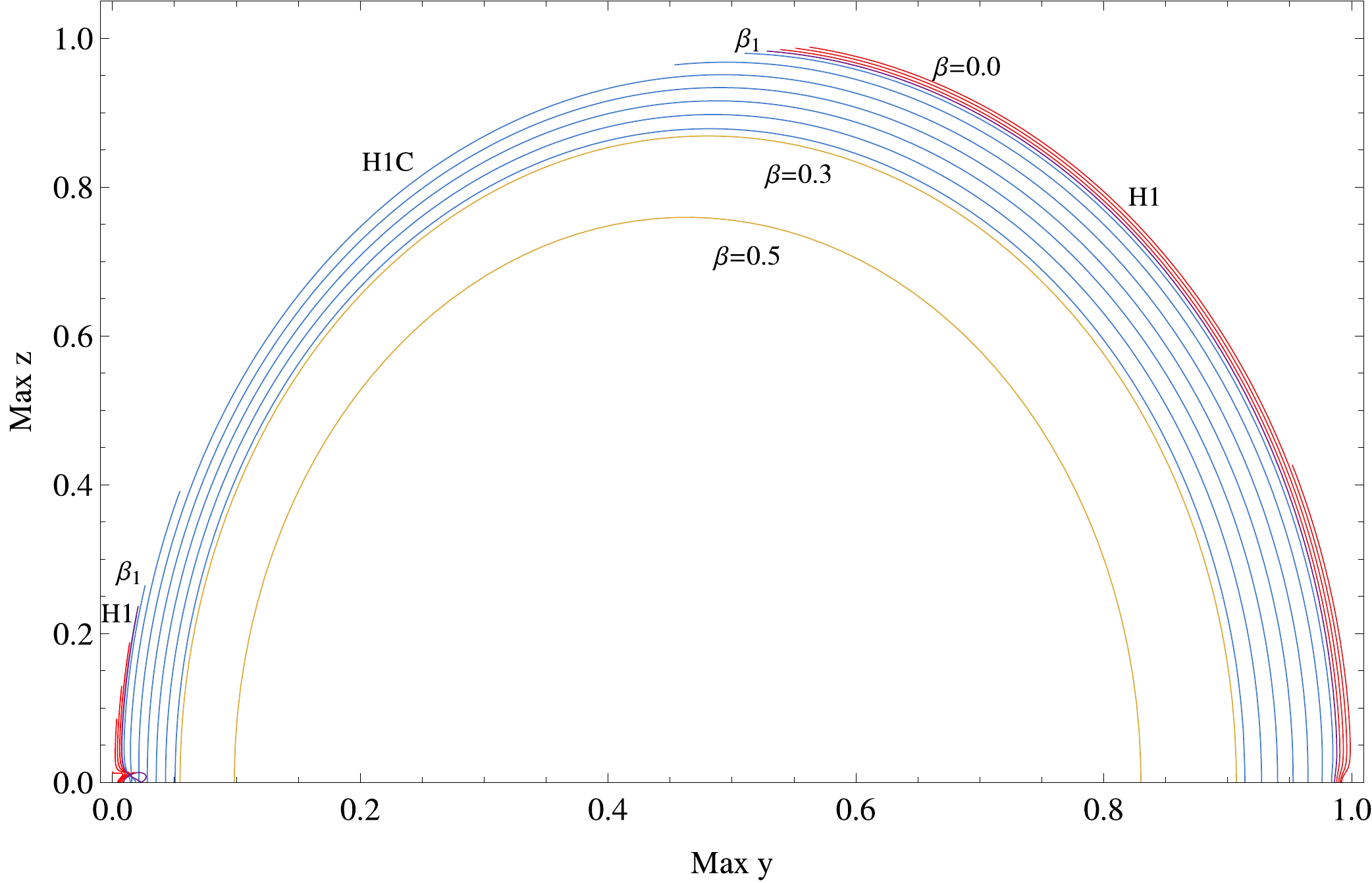}
\caption{\label{fig:bftracks3}
Evolution of H1 and H1C in $\beta$. Each line represents one family. The H1 branch from the L1 family is on the left side of the plot (as shown in Fig. \ref{fig:bftracks1}), and the branch from the C2 family on the right. Red lines are used for values of $\beta < \beta_1$ (i.e. H1), blue for $\beta_1<\beta<\beta_2$ and yellow for $\beta>\beta_2$ (i.e. both blue and yellow lines are H1C). The critical case at $\beta=\beta_1$ is shown in purple as before. H1 is shown for $\beta=(0,0.01,0.02,0.03)$ and H1C for $\beta=(0.04,0.05,0.08,0.12,0.16,0.2,0.24,0.28,0.3,0.5)$
}
\end{figure}

Fig. \ref{fig:bftracks1} shows how the H1 family (in red) is split into the two families H1B and H1C (both in blue) for values of $\beta$ up to $0.05$. Fig. \ref{fig:bftracks3} shows H1C and its eventual connection to the family from C2, while Fig. \ref{fig:bftracks2} shows H1B over a larger range of $\beta$. Examples of the H1B and H1C families are shown in Figures \ref{fig:H1Corbit_04}, \ref{fig:H1Borbit_039}, \ref{fig:H1Borbit_289} and \ref{fig:H1Corbit_5} in the Appendix. As $\beta$ increases the H1B families increase in amplitude, and move closer to the Sun. The H1B family is dramatically different from the classical Earth-Sun case, although it is similar in shape to H1 at $\mu=0.5$ in the classical CRTBP. The H1B family is the family seen in \citet{McInnes2000} from the L1 branch point near $\beta=0.055$. 

\begin{figure}[bp]
\centering
\includegraphics[width=0.8\textwidth]{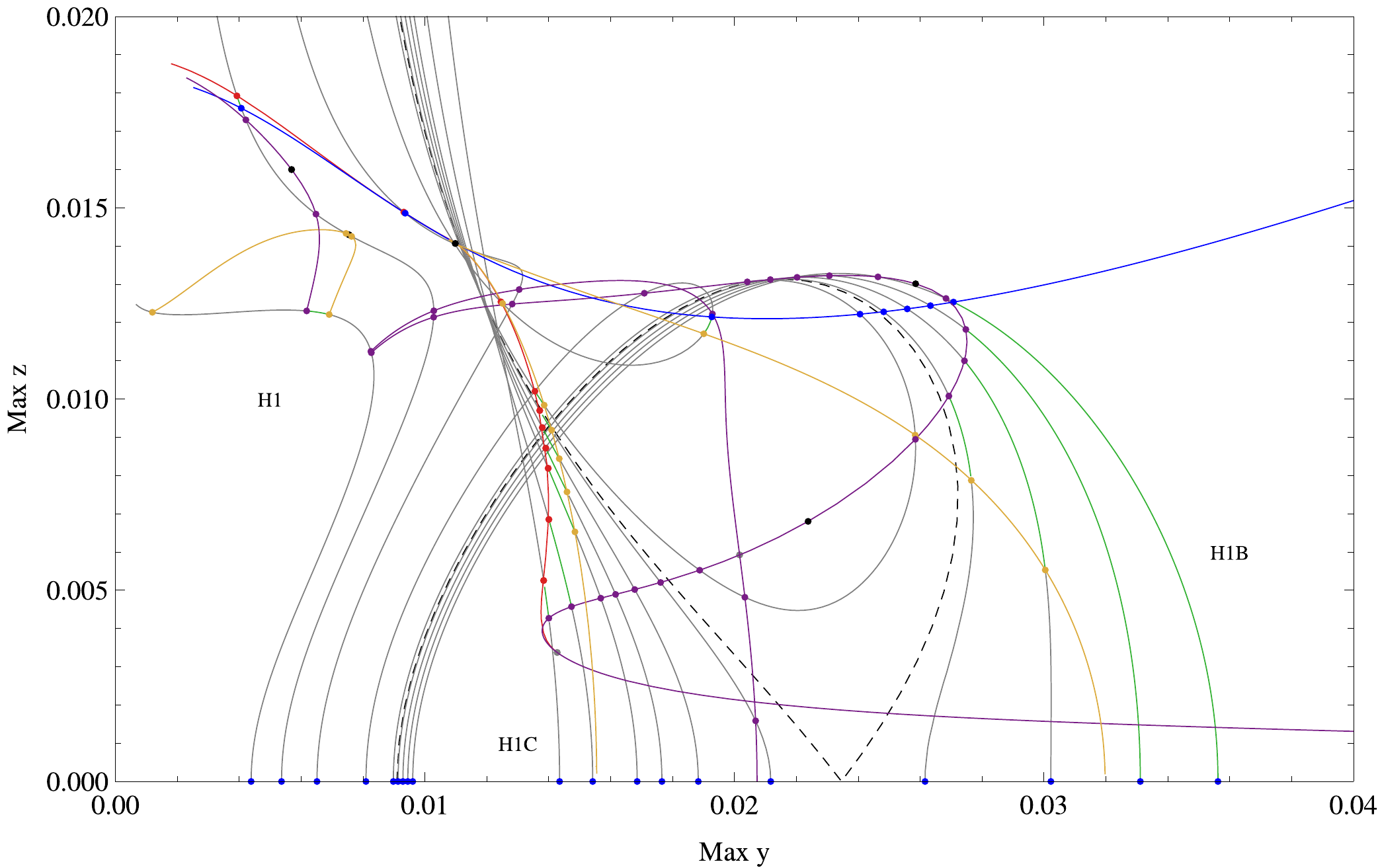}
\caption{\label{fig:stabdata}Stability evolution of the north branch of the H1, H1B and H1C families as the lightness number $\beta$ increases. Families are represented by the periodic orbits maximum $y$ and maximum $z$ values as before (c.f. Fig. \ref{fig:bftracks1}). The green portion of each family indicates the order-0 instability region. The loci of folds are shown in yellow, period-doubling bifurcations in purple, branch points in blue, and the Krein collision in red. The position of folds in $\beta$ on these loci are shown as black points. H1 is shown for $\beta=(0.0,0.01,0.02,0.032,0.038)$, H1B for $\beta=(0.039,0.04,0.041,0.042)$ and H1C for $\beta=(0.039,0.04,0.041,0.042,0.045,0.05)$. The critical family between H1 and H1B and H1C at $\beta=\beta_1$ is shown as a dashed line. The points at which the various loci intersect the plotted families are marked. The locations where two of the period-doubling bifurcations collide and where the Krein collision meets a period-doubling bifurcation are also marked with gray points
}
\end{figure}

As $\beta\to\beta_2$ the H1B and HR branch points in the RS family move together and eventually merge, as shown in panel 4 of Fig. \ref{fig:bfdiag}. As a result, at $\beta=\beta_2$ the H1B and HR families merge into one, labeled H1R. Note that, as we have defined the families, it is the north branch of the H1B family that merges with the south branch of the HR family to form the south branch of H1R. For $\beta>\beta_2$ this new H1R family starts at the L1 branch point and now no longer connects to the RS family, but ends in collision with $m_1$. An example of the H1R family is shown in Fig \ref{fig:H1Rorbit_3} in the Appendix. The configuration of families in this case is shown in panel 5 of Fig. \ref{fig:bfdiag}. Fig. \ref{fig:bftracks2} shows in more detail how the H1B and H1R families approach each other and at $\beta=\beta_2$ merge together (shown in green). The subsequent evolution of H1R is then shown in yellow. The evolution of H1C for $\beta>\beta_2$ is shown in Fig. \ref{fig:bftracks3} in yellow, where it can be seen that the family remains largely unchanged from the $\beta_1 < \beta < \beta_2$ region. In contrast to H1, H1B and H1R the amplitudes of the periodic orbits of H1C decrease as the sail lightness number $\beta$ increases. 

Although \citet{Farres2010} calculate the halo family for $\beta=0.051689$, which is in the H1B regime, the family is only shown near the branch point from the L1 family so the connection to the RS family is not apparent. The linear stability and shape of the family is however consistent with the results of this work. While \citet{Farres2010} also see the merger of different branches of families (the north and south branches of H1 with L1) in the non-Hamiltonian case with the sail angle $\delta\neq 0$ this is a different phenomenon to that observed here. In the case $\delta \neq 0$ the spatial symmetry of the CRTBP is destroyed and the symmetry breaking pitchfork bifurcation from L1 to H1 can no longer exist. Here, in the RSCRTBP, no symmetries are lost as $\beta$ is changed, but families with the same symmetries merge as two individual symmetry breaking pitchfork bifurcations collide. This difference demonstrates the wide range of dynamical phenomena that can be generated by a model of a solar sail.

\section{Linear stability}
\label{sec:stability}

Section \ref{sec:numcont} described how information provided by the Floquet multipliers and continuation of folds, branch points, period-doubling bifurcations and Krein collisions can be used to map out regions of linear stability. We use this to investigate the evolution of the linear stability of the H1, HR, H1B, H1C and H1R families. The results are included in the topological bifurcation diagram in Fig. \ref{fig:stabdiag}, and included as color-coding of the family branches in Fig. \ref{fig:stabdata} (which shows the same branches as Fig. \ref{fig:bftracks1}). The loci of branch points, period-doubling bifurcations and folds shown in Fig. \ref{fig:stabdata} are continuation data. The locus of the Krein collision has been computed in part using the extended system presented in Sec. \ref{subsec:krein} and in part by tracking sign changes of the quantity $B-A^2/4-2$ on the solution surface of the system (\ref{eq:bvp1})--(\ref{eq:bvp3}) defining the periodic orbits. (For small $\beta$ the Krein collision orbit requires increasingly fine meshes making the extended system computationally expensive.) As inverse Krein collisions are not bifurcations and do not indicate a change of linear stability they have not been tracked. However, their location is marked in the topological diagram (Fig. \ref{fig:stabdiag}) to clarify the evolution of the Floquet multipliers of the H1 and H1C families. 

At low values of $\beta$ the linear stability changes rapidly, with small regions of order-0 instability appearing and disappearing in H1. The H1B and H1R families are largely either of order-1 or order-0, while H1C is mostly of order-2. As already discussed, the H1 family in the classical case is known to have a small region of order-0 instability bounded by a fold and period-doubling bifurcation for the Earth-Sun mass ratio \citep{Howell1984}, as shown in Fig. \ref{fig:H1orbit_0} in the Appendix. This classical family is shown on the left side of Fig. \ref{fig:stabdiag}. As $\beta$ increases to just above 0.01 the fold and period-doubling bifurcation disappear as they collide and annihilate with another fold and period-doubling bifurcation respectively. The loci of each of these bifurcations are shown in Fig. \ref{fig:stabdata}. 

At $\beta=0.01$ the periodic orbits that make up the H1 family have increased in amplitude sufficiently so that the \bp{H1}{W5} branch point reappears. A region of order-0 instability is bounded by \bp{H1}{W5} and a Krein collision. Although this region appears fairly large in the topological diagram in Fig. \ref{fig:stabdiag} the continuation data in Fig. \ref{fig:stabdata} shows that it is much smaller in physical and parameter space. At $\beta\approx0.0209$ another pair of folds appears, splitting the order-0 region into two, one part of which disappears completely after the W5 branch point \bp{H1}{W5} crosses the period-doubling bifurcation along the family. At the crossing point the orbit has a quadruple Floquet multiplier $+1$ and a double Floquet multiplier $-1$, a codimension-two phenomenon. Such bifurcations, though rare, are to be expected as two parameters ($\beta$ and the energy level) are being varied simultaneously here. 

The two period-doubling bifurcations of the original H1 family undergo many changes after this point, folding several times with respect to  $\beta$. At $\beta\approx0.0374$ two merge and briefly disappear before reappearing. This is where the two lines appear to cross between the $\beta=0.0372$ and $\beta=0.038$ families in Fig. \ref{fig:stabdiag}.

The appearance of \bp{RS}{H1B} and \bp{RS}{H1C} in RS, and how they split the H1 family into H1B and H1C, is shown topologically in Fig. \ref{fig:stabdiag}. Slightly before $\beta_1$ a period-doubling bifurcation and a fold in H1 cross to create another order-0 region, which rapidly increases in size as it evolves into H1B. The fold disappears, as does the period-doubling bifurcation, leaving a large order-0 region in H1B bounded by \bp{H1B}{W5} and \bp{RS}{H1B}, as shown in Fig. \ref{fig:stabdata}. There is no change in the linear stability of the family after $\beta\approx0.04$. The linear stability regions mapped out here for H1 and H1B also agree with the linear stability of the individual H1 and H1B families between $\beta=0$ and $\beta=0.06$ presented in \citet{McInnes2000}. 

The H1C family has a small order-0 region bounded by a fold, and subsequently a period-doubling bifurcation, and a Krein collision. This region increases in size, as shown in Fig. \ref{fig:stabdata}, until $\beta\approx0.067$, when the Krein collision collides with the period-doubling bifurcation (i.e. the periodic orbit has a quadruple Floquet multiplier $-1$) and moves off onto the real axis and becomes an inverse Krein collision. 

The linear stability of the part of H1 that branches off from the C2 family does not appear to change with $\beta$; the period-doubling bifurcation and inverse Krein collision are present for values of $\beta$ up until at least 0.5. At $\beta\approx0.16$ the collision with $m_2$ disappears such that this part of H1 merges with H1C. The linear stability of H1C then does not change up to at least $\beta=0.5$, although the first period-doubling bifurcation does gradually approach the RS branch point.

There are no bifurcations in the HR family and it is always entirely of order-0. When it merges with H1B to form H1R the linear stability of both families persists, and H1R is of order-0 after the branch point to W5 until the end of the family in the collision with $m_1$. 

The linear stability regions and bifurcations in H1, H1B, H1C and H1R are also shown colour-coded in the example families plotting in Figs. \ref{fig:egorbits1} and \ref{fig:egorbits2} in the Appendix.

\section{Conclusions}
\label{sec:conclusion}

We have used the boundary-value problem numerical continuation methods in AUTO \citep{Doedel1991,Doedel2011} with the unfolding method of \citet{Munoz2003} to investigate how the H1 halo family evolves in the Earth-Sun RSCRTBP as the sail lightness number $\beta$ (or radiation pressure) increases. We confirm the results of \citet{McInnes2000} that a significant change in the shape of the H1 family occurs at low values of $\beta$. We also find that the change is linked to a bifurcation in the retrograde satellite family RS about the Earth at $\beta=\beta_1\approx0.0387$ which results in the appearance of two additional branch points that split the original halo family H1 into the two families H1B and H1C. One of these new branch points collides with an existing branch point in RS at $\beta=\beta_2\approx0.289$, further changing the nature of the halo family by causing the H1B and HR families to merge, resulting in the appearance of the H1R family. We note that the range of $\beta$ values given by \citet{sunjammer} for the upcoming Sunjammer mission places that sail within required range for the H1B family.  

The continuation of the RS and C2 families in $\beta$ shows that the two are a single family in the $(\beta-C^\prime_J)$ parameter plane. That these two families connect at $\beta=1$ is not surprising, and indeed similar connections between families occur in the classical CRTBP for limiting values of the mass ratio such as $\mu=0$ and $\mu=0.5$. The connection of H1 to HR at $\beta=\beta_2$ is perhaps less predictable, and indicates that the RSCRTBP could also be used as a novel means of generating some classical families. For example, continuing H1 in $\beta$ to obtain either the H1R family or H1B family (and thus the RS family) provides a very quick method of generating the classical HR and RS families through continuation back to $\beta=0$.

A consequence of the additional branch points in the RS family between $\beta_1$ and $\beta_2$ is that a region of the retrograde satellite family near the Earth is not of order-0 instability, and thus will possess stable and unstable manifolds. This is in contrast to the classical retrograde satellite family that has order-0 instability from its start to the branch point with the HR family at very large amplitude. The existence of such manifolds may lead to the possibility of low-energy transfers to retrograde orbits for a solar sail that are not obtainable with traditional spacecraft.

The linear stability of the H1, H1B and H1C families in the RSCRTBP changes rapidly below $\beta\approx0.04$, with several regions of order-0 instability appearing and disappearing in H1. The families H1B and H1R have large regions of order-0 instability, while H1C has none above $\beta\approx 0.067$, and is in general very unstable. While it might be expected that as $\beta$ increases the system becomes similar to that of the CRTBP with a higher mass ratio the results presented here show that the effects of radiation pressure are more subtle than that.

The detailed results presented here for the \lone halo family in the Earth-Sun RSCRTBP demonstrate that even at realistically low values of $\beta$ the families of periodic orbits obtainable by solar sails show significant differences to the classical case.

\begin{acknowledgements}
  P.V. acknowledges funding from the University of Portsmouth. The
    research of J.S.\ is supported by EPSRC Grant EP/J010820/1.
\end{acknowledgements}


\section*{Appendix}

\begin{figure}[!ht]
\centering
\subfigure[\label{fig:l1eseg}$\beta=0$]{\includegraphics[width=0.32\textwidth]{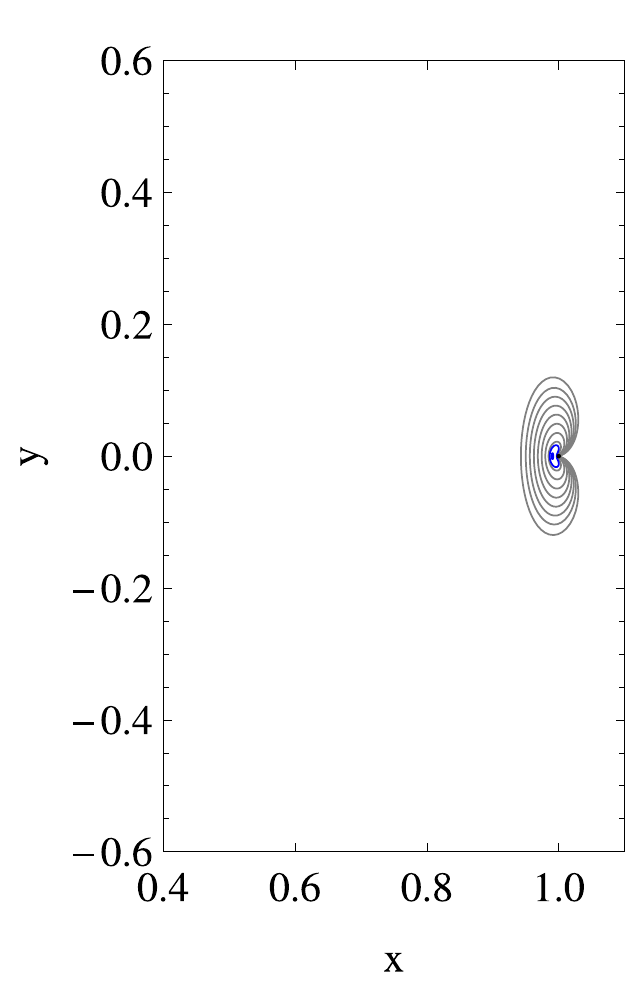}}
\subfigure[$\beta=0.05$]{\includegraphics[width=0.32\textwidth]{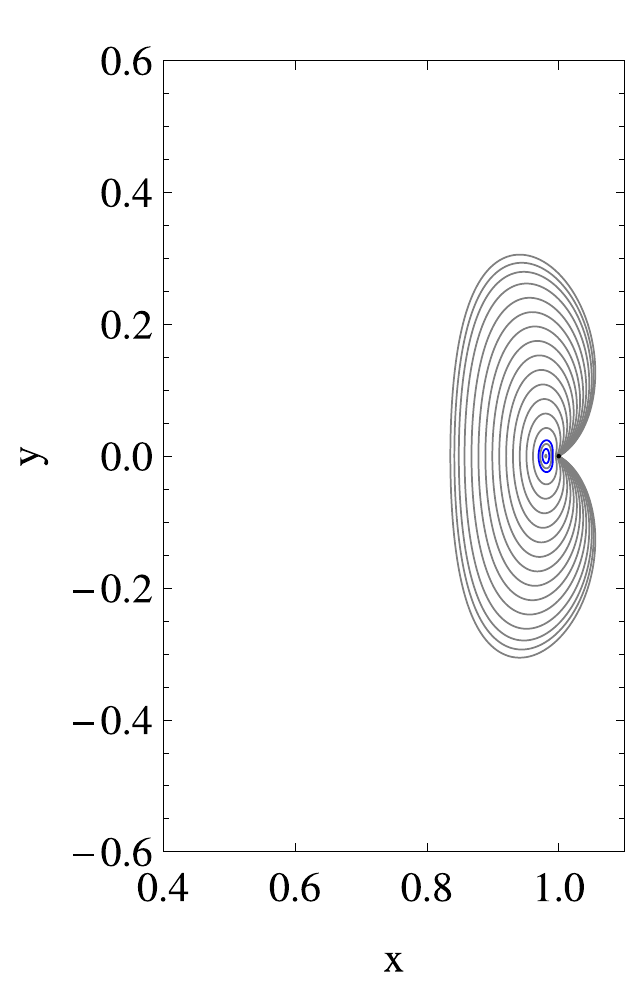}}
\subfigure[$\beta=0.5$]{\includegraphics[width=0.32\textwidth]{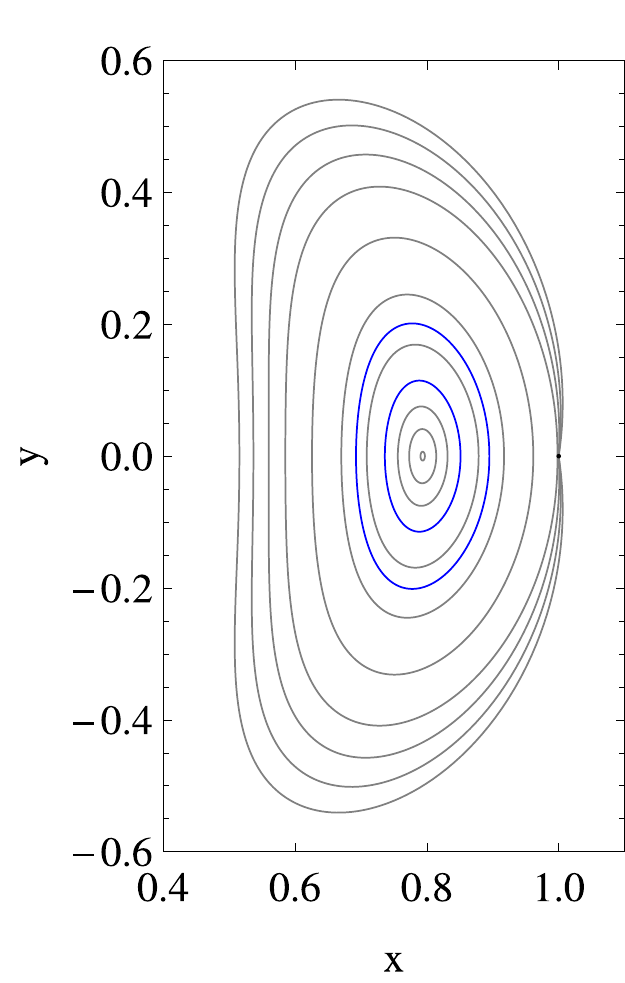}}
\caption{\label{fig:l1ev} Examples of the L1 family for different values of $\beta$ in the Earth-Sun RSCRTBP, with branch points marked in blue. The location of the Earth ($m_2$) is shown as a black dot}
\end{figure}

\begin{figure}[!ht]
\centering
\subfigure[\label{fig:c2eseg}$\beta=0$]{\includegraphics[width=0.32\textwidth]{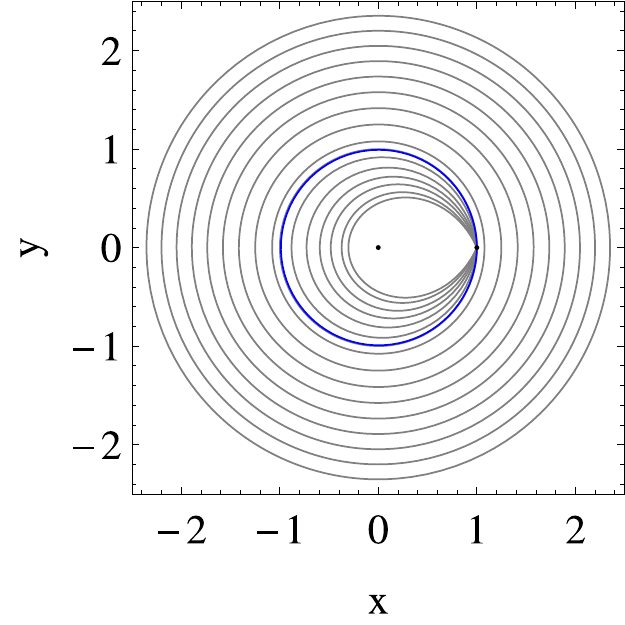}}
\subfigure[$\beta=0.05$]{\includegraphics[width=0.32\textwidth]{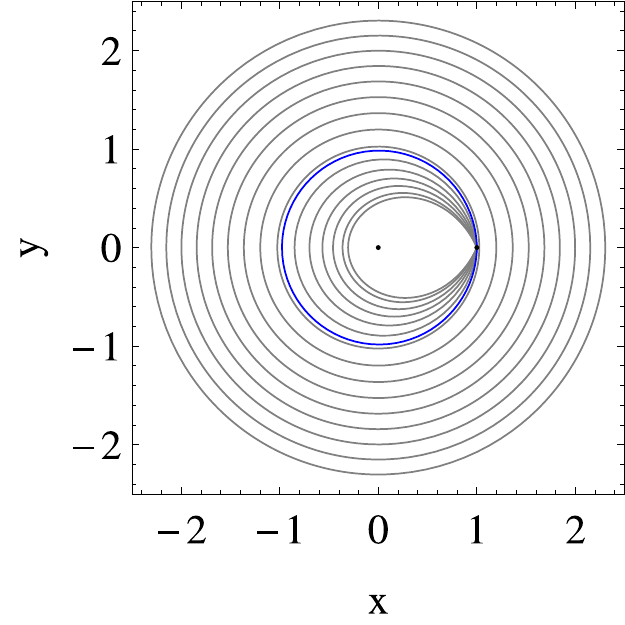}}
\subfigure[$\beta=0.5$]{\includegraphics[width=0.32\textwidth]{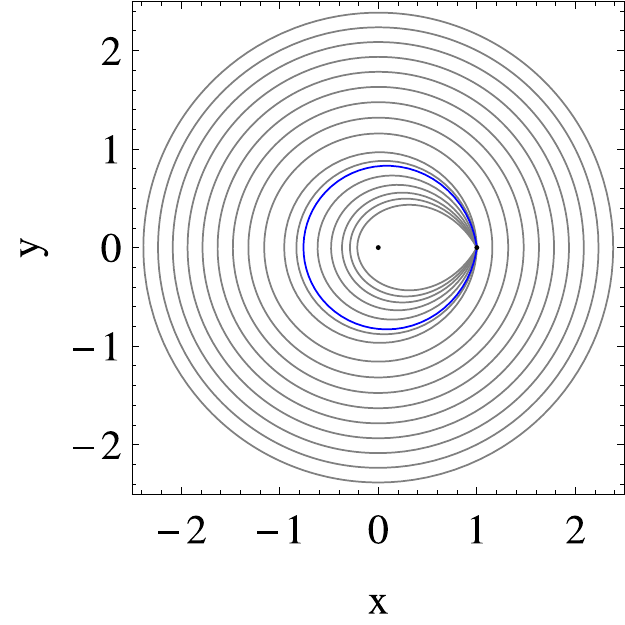}}
\caption{\label{fig:c2ev} Examples of the C2 family for different values of $\beta$ in the Earth-Sun RSCRTBP, with branch points marked in blue. The location of the Earth ($m_2$) and Sun ($m_1$) are shown as black dots. The families continue out to infinity in all cases. The C2 family remains relatively unchanged as the sail lightness number increases}
\end{figure}

\begin{figure}[!ht]
\centering
\subfigure[\label{fig:rseseg}$\beta=0$]{\includegraphics[width=0.32\textwidth]{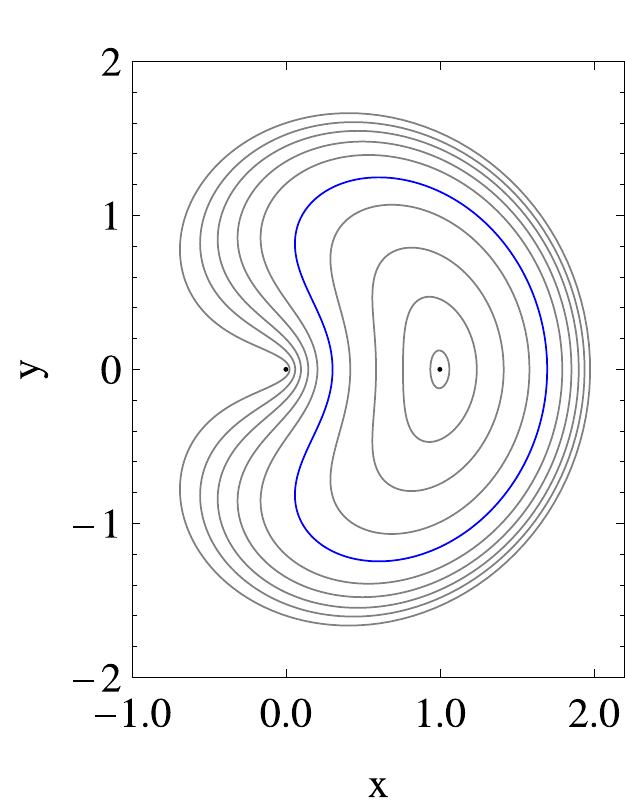}}
\subfigure[$\beta=0.05$]{\includegraphics[width=0.32\textwidth]{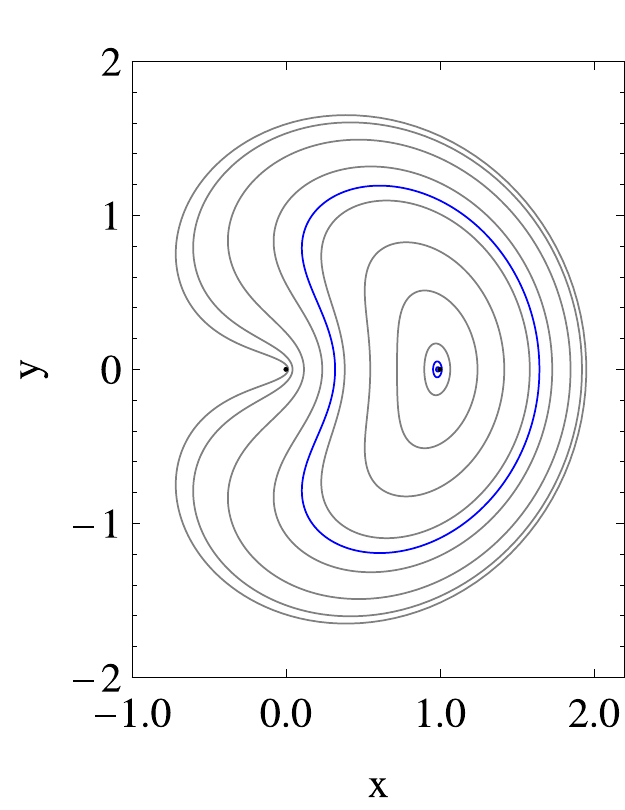}}
\subfigure[$\beta=0.5$]{\includegraphics[width=0.32\textwidth]{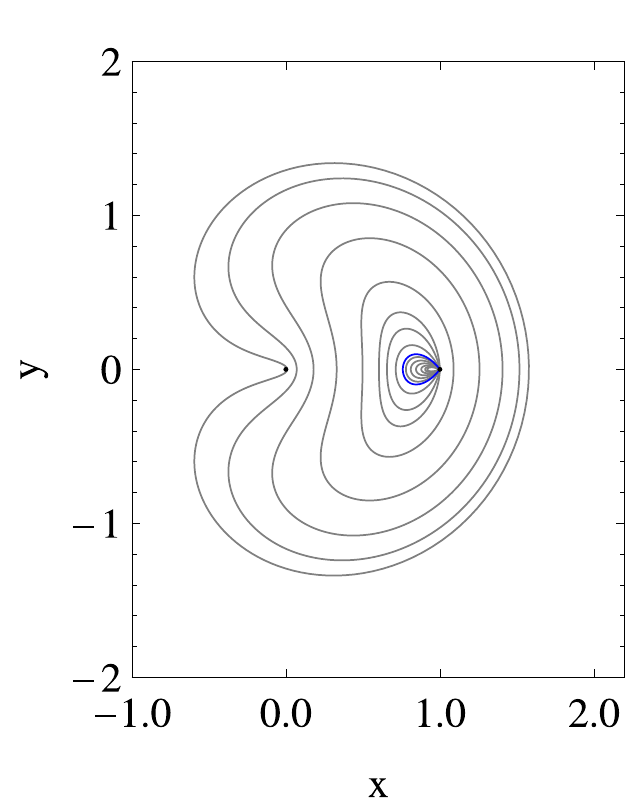}}\\
\subfigure[$\beta=0.05$]{\includegraphics[width=0.32\textwidth]{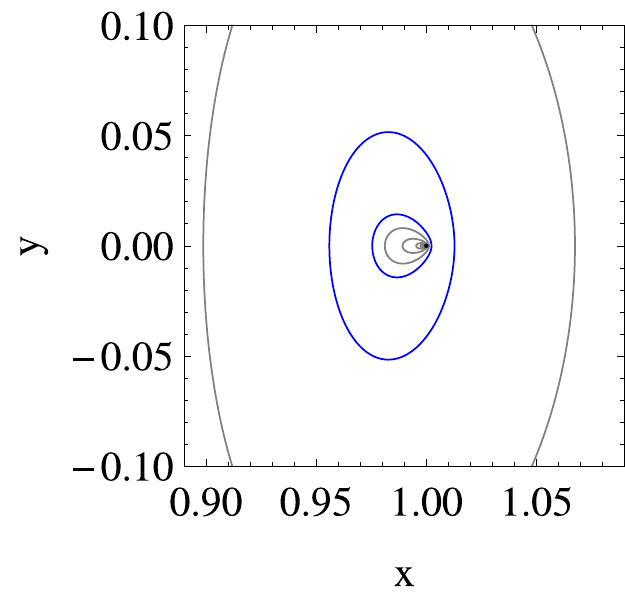}}
\caption{\label{fig:rsev}  Examples of the RS family for different values of $\beta$, with branch points marked in blue in the Earth-Sun RSCRTBP. The location of the Earth ($m_2$) and Sun ($m_1$) are shown as black dots. Each family ends in a collision with the Sun. A close up of the two extra branch points near the Earth is also shown for the $\beta=0.05$ case}
\end{figure}

\begin{figure}[!ht]
\centering
\subfigure[\label{fig:H1orbit_0}H1, north branch from L1 for $\beta=0.0$]{\includegraphics[width=0.49\textwidth]{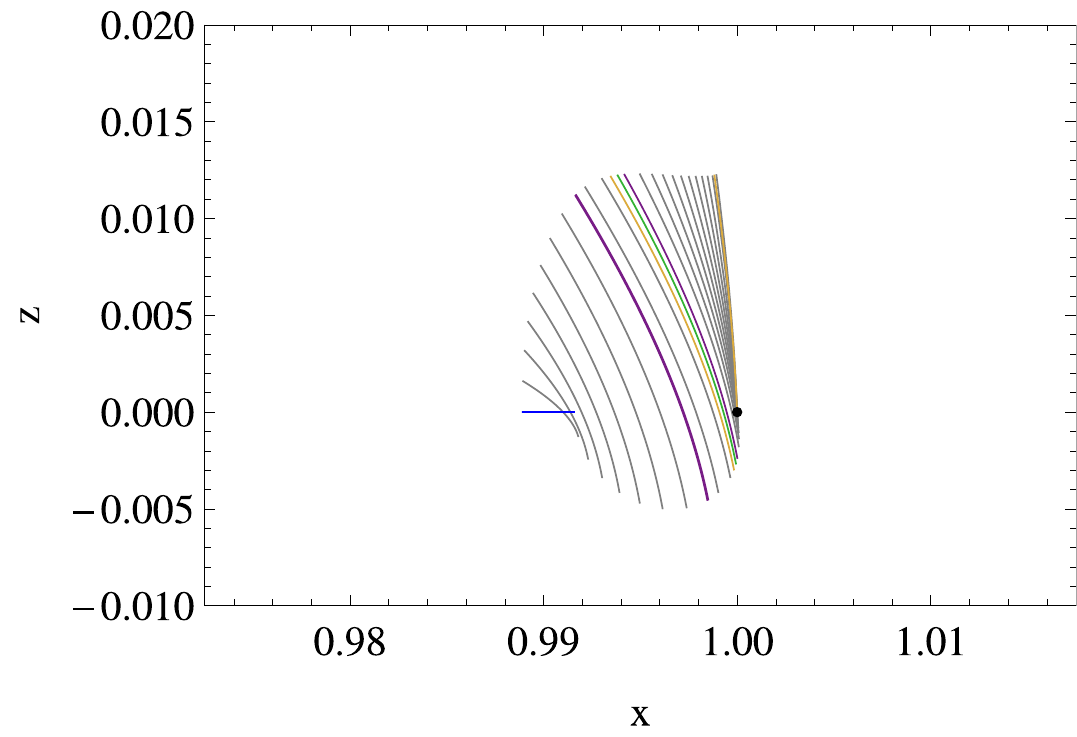}}
\subfigure[\label{fig:H1orbit_1}H1, north branch from C2 for $\beta=0.0$]{\includegraphics[width=0.49\textwidth]{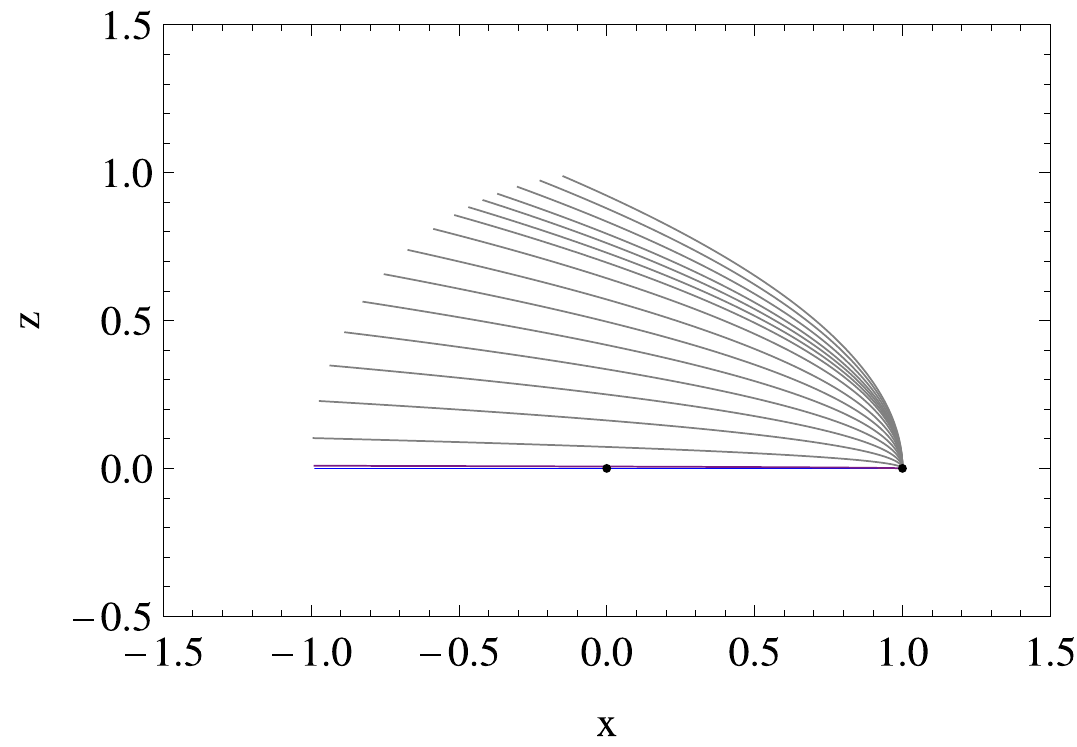}}\\
\subfigure[\label{fig:hreseg}HR, south branch for $\beta=0.0$]{\includegraphics[width=0.49\textwidth]{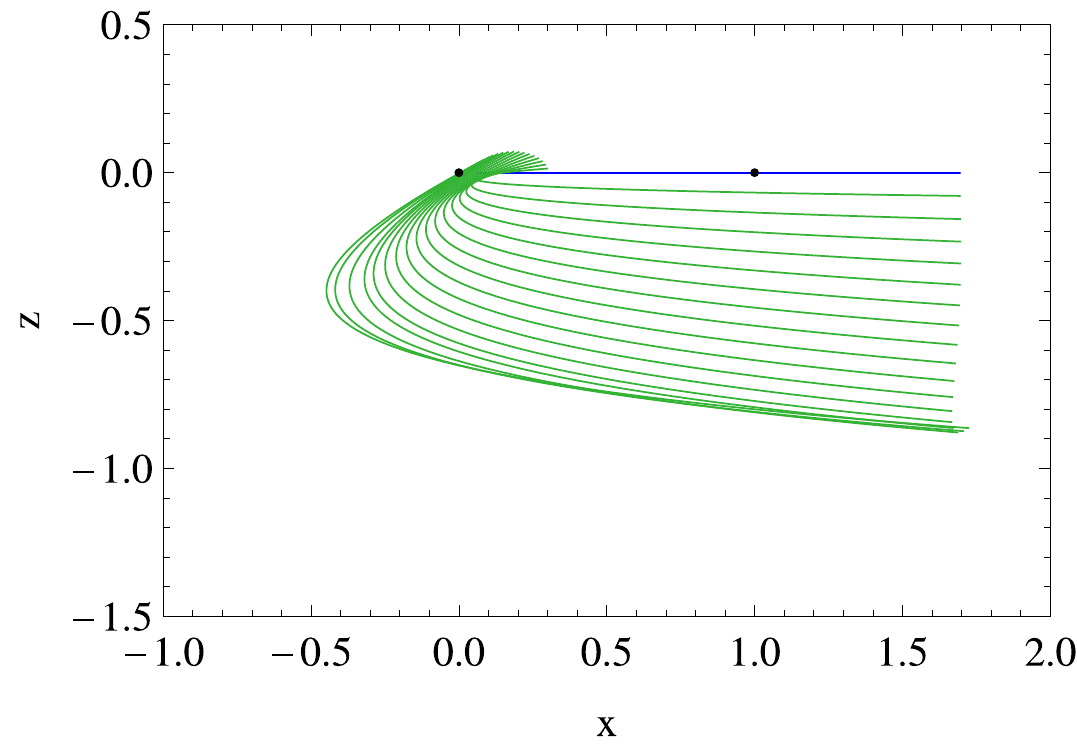}}
\caption{\label{fig:egorbits1} The north branch of the H1 family in the Earth-Sun CRTBP (i.e. $\beta=0$) from (a) the L1 branch point and (b) the C2 branch point, and (c) the south branch of the HR family, shown in the $x-z$ plane. Order-0 instability orbits are shown in green, others in gray. Branching orbits are shown in blue, period-doubling bifurcations in purple and folds in yellow. The positions of the Earth and Sun are also marked. Note that different scales are used in each plot}
\end{figure}

\begin{figure}[p]
\centering
\subfigure[\label{fig:H1orbit_03}H1, north branch from L1 for $\beta=0.03$]{\includegraphics[width=0.49\textwidth]{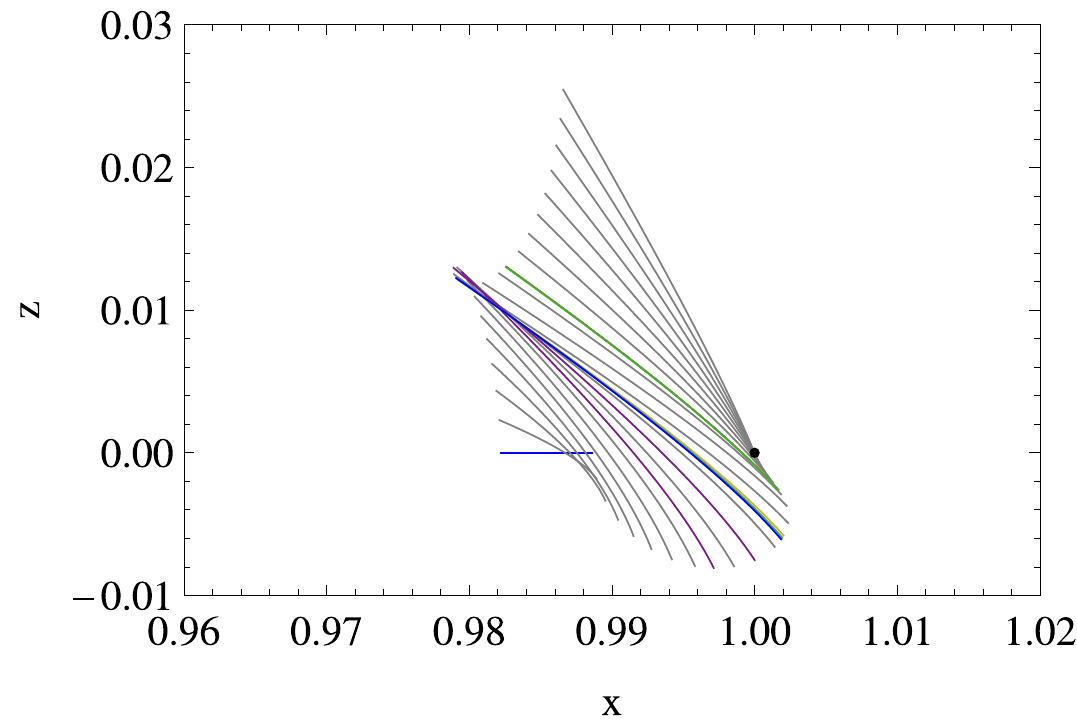}}
\subfigure[\label{fig:H1Corbit_04}H1C, north branch from RS for $\beta=0.04$]{\includegraphics[width=0.49\textwidth]{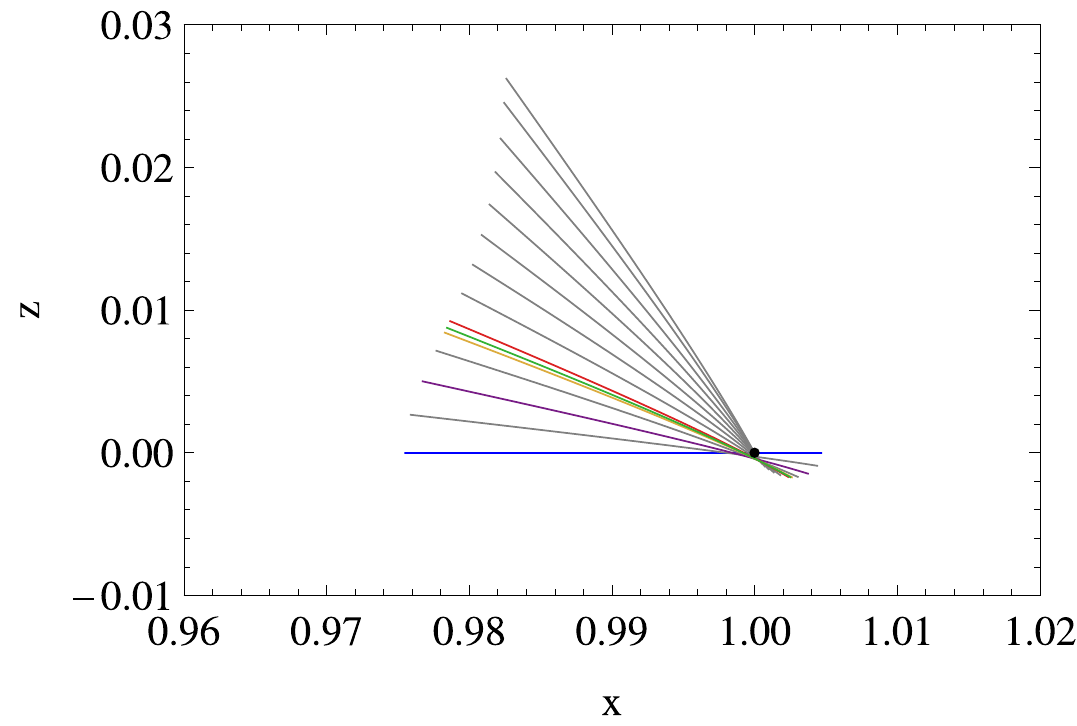}}\\
\subfigure[\label{fig:H1Borbit_039}H1B, north branch for $\beta=0.039$]{\includegraphics[width=0.49\textwidth]{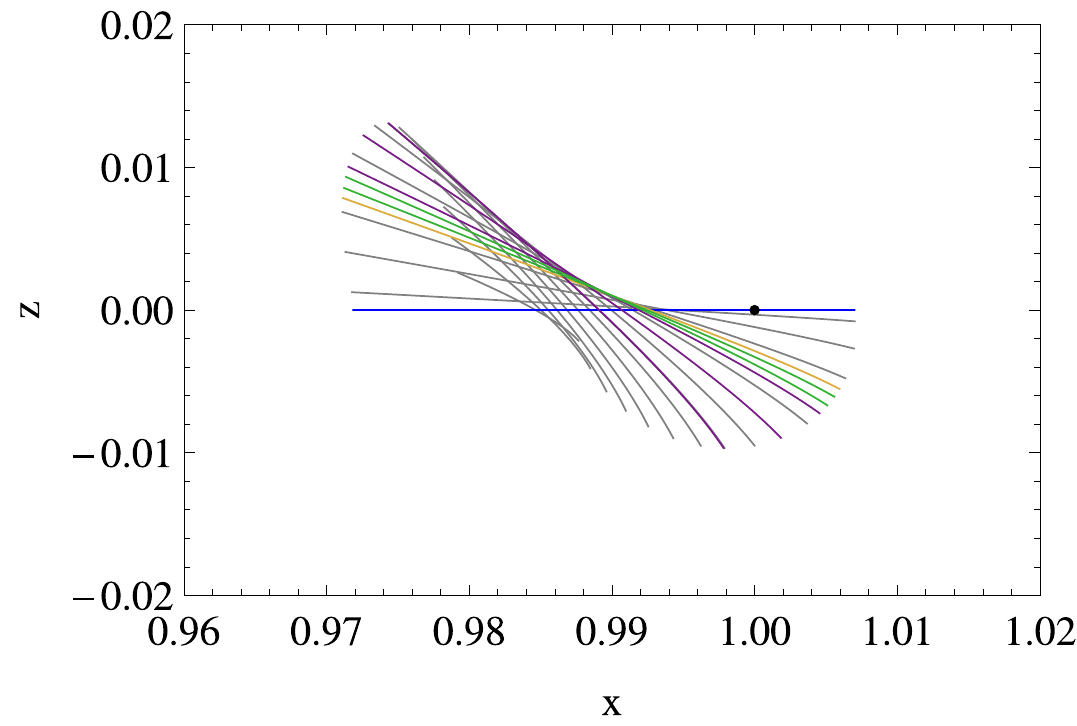}}
\subfigure[\label{fig:H1Borbit_289}H1B, north branch for $\beta=0.289$]{\includegraphics[width=0.49\textwidth]{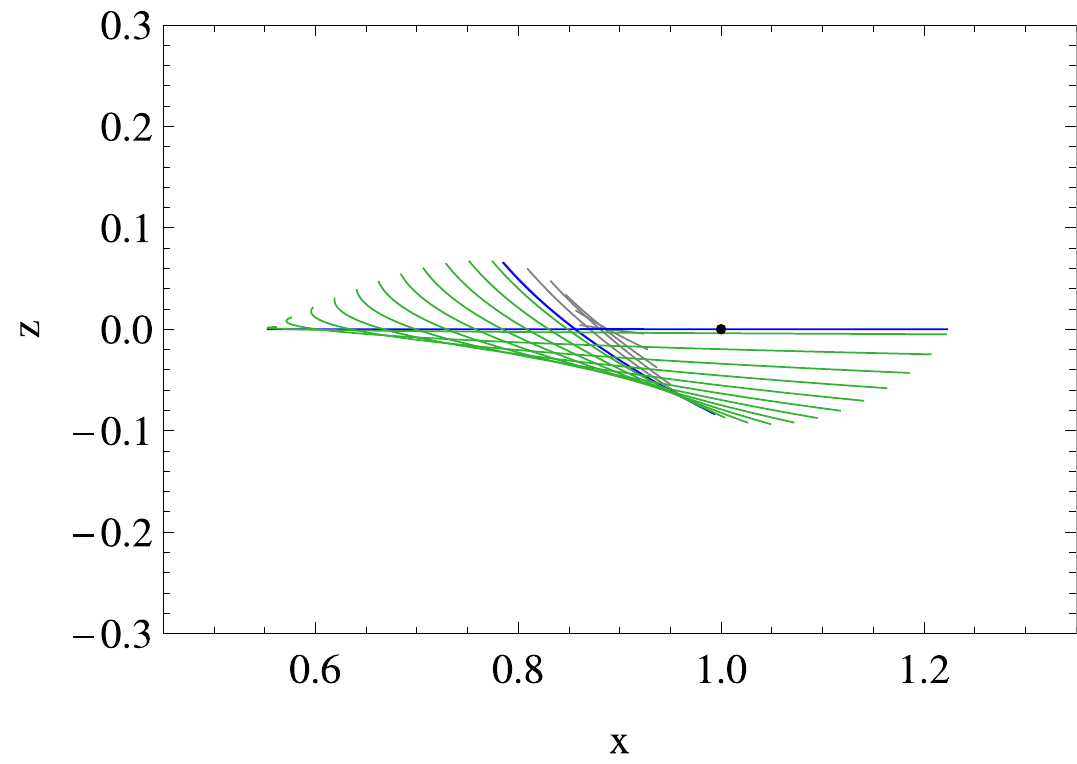}}
\subfigure[\label{fig:H1Corbit_5}H1C, north branch for $\beta=0.5$]{\includegraphics[width=0.49\textwidth]{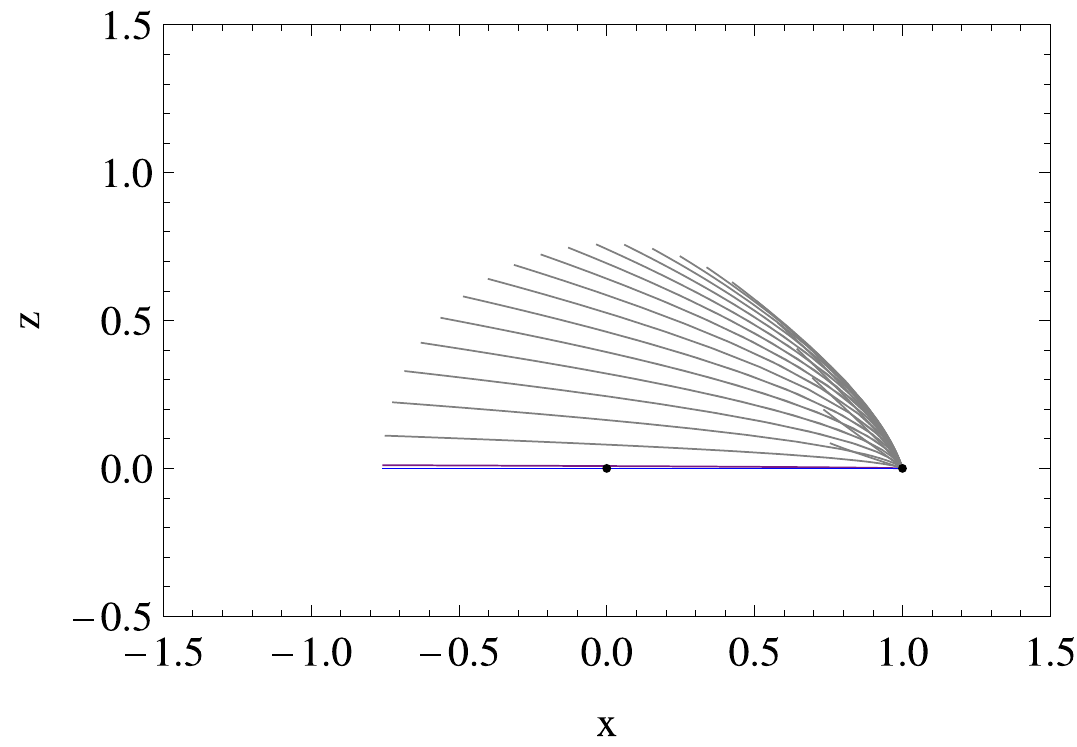}}
\subfigure[\label{fig:H1Rorbit_3}H1R, south branch for $\beta=0.3$]{\includegraphics[width=0.49\textwidth]{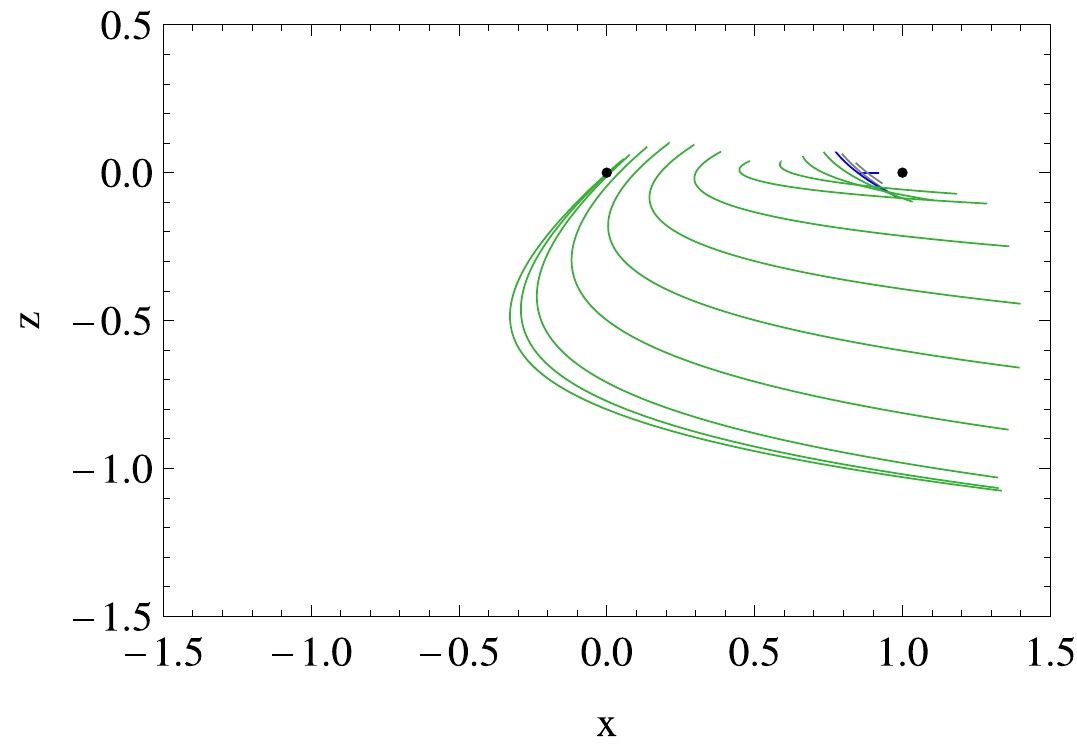}}
\caption{\label{fig:egorbits2}Examples of the H1, H1B, H1C and H1R families shown in the $x-z$ plane for various values of $\beta$ in the Earth-Sun RSCRTBP. Order-0 instability orbits are shown in green, others in gray. Branching orbits are shown in blue, period-doubling bifurcations in purple, folds in yellow and the approximate locations of Krein collisions in red. The positions of the Earth and Sun are also marked. Note that different scales are used in each plot}
\end{figure}

\begin{figure}[!hb]
\centering
\includegraphics[width=0.49\textwidth]{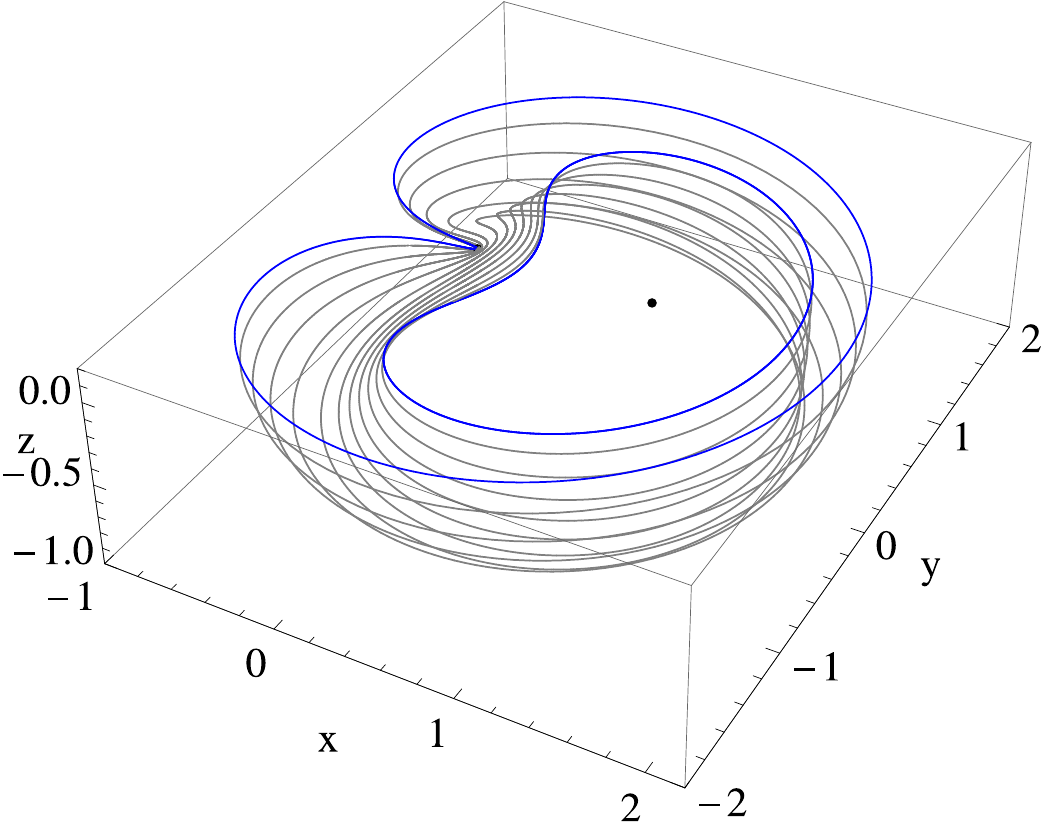}
\caption{\label{fig:hrem}The southern branch of the HR family in the classical CRTBP for the Earth-Moon mass ratio}
\end{figure}

In this appendix we present examples of the families discussed in Sections \ref{sec:periodicorbits} to \ref{sec:stability}. Examples of the L1, C2 and RS families for several values of $\beta$ are shown in Figs. \ref{fig:l1ev}, \ref{fig:c2ev} and \ref{fig:rsev}. Examples of the H1, HR, H1B, H1C and H1R families for the Earth-Sun mass ratio in the RSCRTBP are shown in Figs. \ref{fig:egorbits1} and \ref{fig:egorbits2}. A three-dimensional plot of the HR family for the Earth-Moon CRTBP ($\mu=0.01215$, $\beta=0$) is shown in Fig. \ref{fig:hrem}.

\end{document}